\documentclass[]{spie}  

\usepackage{amsmath,amsfonts,amssymb}
\usepackage{graphicx}
\usepackage[colorlinks=true, allcolors=blue]{hyperref}

\DeclareMathOperator*{\argmin}{argmin}

\usepackage{graphicx}
\usepackage{subcaption}
\usepackage[export]{adjustbox}
\usepackage{wrapfig}

\usepackage{float}

\title{Accurately Measuring Hyperspectral Imaging Distortion in Grating Spectrographs Using a Clustering Algorithm}

\author[a,b]{Matthew C. H. Leung}
\author[a]{Shaojie Chen}
\author[c]{Colby Jurgenson}
\affil[a]{Dunlap Institute for Astronomy and Astrophysics, University of Toronto, 50 St. George St, Toronto, ON, M5S 3H4, Canada}
\affil[b]{Division of Engineering Science, University of Toronto, 40 St. George St, Toronto, ON, \mbox{M5S 2E4}, Canada}
\affil[c]{Harvard-Smithsonian Center for Astrophysics, 60 Garden St, Cambridge, MA, 02138, USA}

\authorinfo{Further author information: (Send correspondence to M.C.H. Leung or S. Chen)\\M.C.H. Leung: E-mail: matthewchingho.leung@mail.utoronto.ca\\  S. Chen: E-mail: sj.chen@utoronto.ca\\ C. Jurgenson: E-mail: colby.jurgenson@cfa.harvard.edu}

\begin{document} 
\maketitle

\begin{abstract}
Grating-based spectrographs suffer from smile and keystone distortion, which are problematic for hyperspectral data applications. Due to this, spectral lines will appear curved and roughly parabola-shaped. Smile and keystone need to be measured and corrected for accurate spectral and spatial calibration. In this paper, we present a novel method to accurately identify and correct curved spectral lines in an image of a spectrum, using a clustering algorithm we developed specifically for grating spectrographs, inspired by $K$-means clustering. Our algorithm will be used for calibrating a multi-object spectrograph (MOS) based on a digital micromirror device (DMD). For each spectral line in a spectrum image, our algorithm automatically finds the equation of the parabola which models it. Firstly, the positions of spectral peaks are identified by fitting Gaussian functions to the spectrum image. The peaks are then grouped into a given number of parabola-shaped clusters: each peak is iteratively assigned to the nearest parabola-shaped cluster, such that the orthogonal distances from the parabola are minimized. Smile can then be measured from the parabolas, and keystone as well if a marked slit is used. Our method has been verified on real-world data from a long-slit grating spectrograph with sub-pixel error, and on simulated data from a DMD-based MOS. Compared to traditional approaches, our method can measure distortions automatically and accurately while making use of more spectral lines. With a precise model and measurement of distortion, a corrected hyperspectral data cube can be created, which can be applied for real-time data processing.  
\end{abstract}

\keywords{Smile, Keystone, Hyperspectral imaging distortion, Grating spectrograph, Spectrograph calibration, Clustering}


\section{Introduction}\label{sec:intro}

Diffraction gratings are widely used as the primary dispersive element in astronomical spectrographs. However, grating-based spectrographs suffer from aberrations which cause hyperspectral imaging distortion, namely, smile distortion and keystone distortion. As a result of this, spectral lines on the image plane are not perfectly straight, but instead are curved and roughly parabolic in shape \cite{Schroeder2000}. The work in this paper aims to correct for these distortions using an automated and accurate method. Ultimately, our method will be used as a part of a procedure to calibrate a multi-object spectrograph (MOS) which uses a digital micromirror device (DMD) as the programmable slit mask.

\subsection{Smile and Keystone Distortion}
\begin{figure}[ht]
    \centering
    \includegraphics[width=0.56\textwidth]{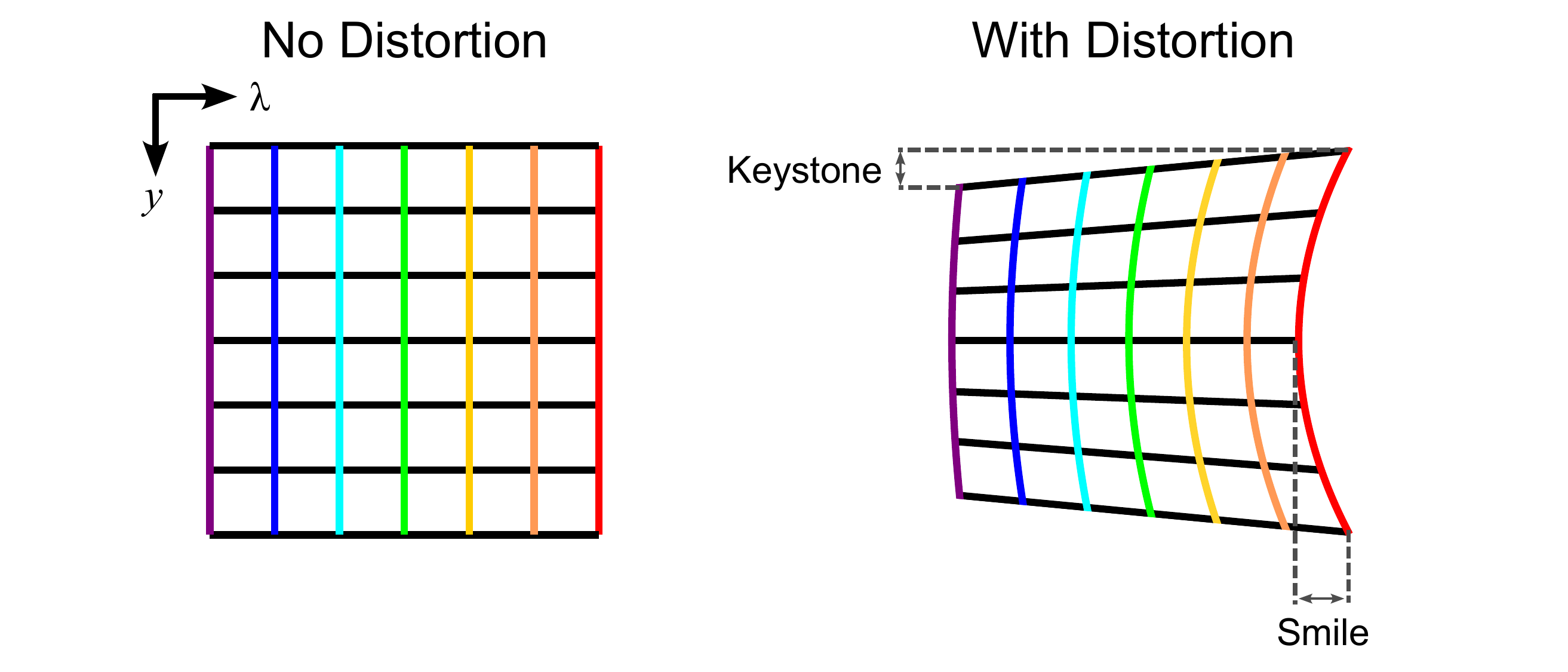}
    \caption{An illustration of the effect of smile and keystone distortion on a spectrum. The horizontal direction is the spectral direction and the vertical direction is the spatial direction.}
    \label{fig:smile_keystone}
\end{figure}

Consider an image of a two-dimensional spectrum from a long-slit grating spectrograph, in which the spatial direction is oriented along the pixel columns and the spectral direction (direction of dispersion) is oriented along the pixel rows, as illustrated in Figure \ref{fig:smile_keystone}. Smile distortion is a change in dispersion angle with respect to field position, causing a shift in wavelength in the spectral domain \cite{Fisher1998,Yokoya2010,Neville2003}. In other words, along a fixed column, the wavelength values will vary. Keystone distortion is a change in magnification with respect to spectral channel, causing inter-band spatial misregistration \cite{Fisher1998,Yokoya2010,Neville2004}. In other words, along a fixed row, the corresponding spatial fields will vary. These distortions are especially problematic for multi-object spectroscopy and hyperspectral imaging applications, in which field positions are off-axis. As a result of this spectral line curvature, the classification of spectral lines is nontrivial.

Suppose that a ray is travelling towards a grating at an off-axis field angle $\gamma$ with respect to the plane of diffraction. $\gamma$ parameterizes the spatial field position. Let $\beta$ be the diffraction angle (dispersion angle). Then according to Ref. \citenum{Schroeder2000}, at a fixed incidence angle and wavelength $\lambda$, the change in $\beta$ with respect to $\gamma$ is:
\begin{equation}
    \frac{d\beta}{d \gamma} = \lambda A \tan{(\gamma)}
\end{equation}
where $A$ is the angular dispersion at $\gamma=0$ (for an on-axis ray). For a given $\gamma$, where $\gamma$ is small, the total change in $\beta$ is:
\begin{equation}\label{eq:delta_beta}
    \Delta \beta = \int_0^{\gamma} \lambda A \tan{(u)} \: du = -\lambda A \ln{(\cos{(\gamma)})} \approx \frac{1}{2} \lambda A \gamma^2 
\end{equation}
where $\frac{1}{2}\gamma^2$ is the Taylor series of $-\ln{(\cos{(\gamma)})}$ about $\gamma=0$, truncated to second degree terms. Hence, for small angles, spectral lines appear roughly parabolic in shape.


\subsection{Traditional Approaches for Distortion Correction}

In traditional approaches for hyperspectral imaging distortion correction, a few spectral lines in the image are selected to use for distortion correction \cite{Neville2003,EsmondeWhite2011}. Firstly, horizontal slices are taken across the image (in the spectral direction), in order to find the positions of the spectral peaks (in pixel coordinates) corresponding to those selected spectral lines. Along each horizontal slice, the positions of the spectral peaks are the positions of the maxima along the slice. These spectral peaks can be selected manually or can be found by using by fitting Gaussian functions to the slice \cite{EsmondeWhite2011}. With the set of spectral peak points belonging to each selected spectral line, and the associated wavelength value for each selected spectral line, corrections can then be made. Since the spectral lines are roughly parabolic in shape, for each spectral line, a quadratic polynomial is fitted to the set of spectral peak points corresponding to that spectral line \cite{Neville2003,EsmondeWhite2011,Hong17}. A relationship between wavelength and the parameters of the quadratic polynomials can then be derived.

However, there are several problems with traditional approaches. Each spectral peak point needs to be assigned to a certain spectral line, and this process could be time consuming or restrictive. In traditional approaches, only a few spectral lines are selected, and oftentimes, the spectral peak points corresponding to these selected spectral lines are first manually selected and then their positions are refined by fitting a Gaussian or sinc function \cite{EsmondeWhite2011}. This is because one cannot simply just take all spectral peaks points in some range of columns to be of the same wavelength, since the spectral lines are curved. This is especially the case if there are several spectral lines which are close together. In this case, when taking a slice across a single column in the image, there can be several spectral lines which pass through the same column. Hence, automatically assigning spectral peak points to their corresponding spectral line may be difficult in this case. Manual intervention may be required to ensure that spectral peak points are correctly assigned to the correct spectral line, and this can be time consuming. Therefore, oftentimes to avoid these problems, only a few spectral lines which are prominent are selected.


\subsection{Approaching Distortion Correction as a Clustering Problem}

The positions of spectral peaks can be obtained in a relatively straightforward manner, by fitting Gaussian functions to horizontal slices of the image. In fact, it is relatively easy to obtain the positions of all spectral peaks in an image, and not just a the spectral peaks corresponding to a few spectral lines. However, the challenge is in ensuring that each spectral peak point is assigned to the correct spectral line, and hence traditional approaches only select a few prominent spectral lines. If we want to make use of more spectral lines in an image for distortion correction, while also being able to automatically assign spectral peak points to their corresponding spectral lines, one possible approach is to frame this as a clustering problem. Clustering is the unsupervised classification of unlabelled data into groups \cite{Jain1999}. After the positions of all spectral peaks in an image are obtained through Gaussian fitting, these points are considered as unlabelled data because there is generally no definite knowledge of which spectral lines they belong to. Since we want to classify each of these spectral peak points according to wavelength, this fits the description of a clustering problem.

In this paper, we present a novel method to accurately detect and correct curved spectral lines in an image of a spectrum. This is done using a clustering algorithm we developed specifically for long-slit grating spectrographs, inspired by $K$-means clustering \cite{MacQueen1967,Mannor2011}. $K$-means aims to classify a set of points into $K$ disjoint clusters, such that each point belongs to the cluster with the nearest centroid (mean). The algorithm we developed is called $K$-parabolas, and this algorithm aims to classify a set of points into $K$ disjoint parabola-shaped clusters, such that each point belongs to the nearest parabola. Some constraints which are specific to spectra are also imposed. In addition, in $K$-parabolas, the quantity we want to minimize is different but analogous: In $K$-means, we want to minimize the Euclidean distance between each point and its assigned cluster's centroid \cite{Na2010}; In $K$-parabolas, we will minimize the orthogonal distance between each point and its assigned cluster's parabola.


\subsection{Outline of this Paper}

An outline of this paper is presented as follows. Consider an image of a spectrum. The first step is to identify the positions of spectral peaks corresponding to spectral lines in the image; this will be discussed in Section \ref{sec:spec_lines_id}. Once these spectral peak points are obtained, the problem we are trying to solve will be formally defined in Section \ref{sec:problem}. The $K$-parabolas algorithm will then group the spectral peak points into disjoint parabola-shaped clusters; this will be discussed in Section \ref{sec:k_parabolas}. The parabolas representing each cluster in the result of $K$-parabolas will then be refined, and relationships between the parabola parameters and wavelength will be determined, in order to construct a pixel-to-wavelength mapping for the image; this will be discussed in Section \ref{sec:refine_and_map}. 

Our method will be firstly demonstrated and verified using calibration lamp spectrum images taken by a long-slit grating spectrograph currently being developed at the Dunlap Institute for Astronomy and Astrophysics at the University of Toronto, as a pathfinder instrument for multi-object spectroscopy. Our spectrograph has a wavelength range of 400 nm to 900 nm, and its details will be discussed in Appendix \ref{sec:appendix_spec}. Our method will then be extended and demonstrated on simulated data from a DMD-based MOS also currently being developed at the Dunlap Institute. This will be discussed in Section \ref{sec:DMD-MOS}. More information about our DMD-based MOS can be found in another paper published in these proceedings\cite{Chen2022}.


\section{Identifying the Positions of Spectral Peaks}\label{sec:spec_lines_id}

\begin{figure}[H]
    \centering
    \includegraphics[width=0.5\textwidth]{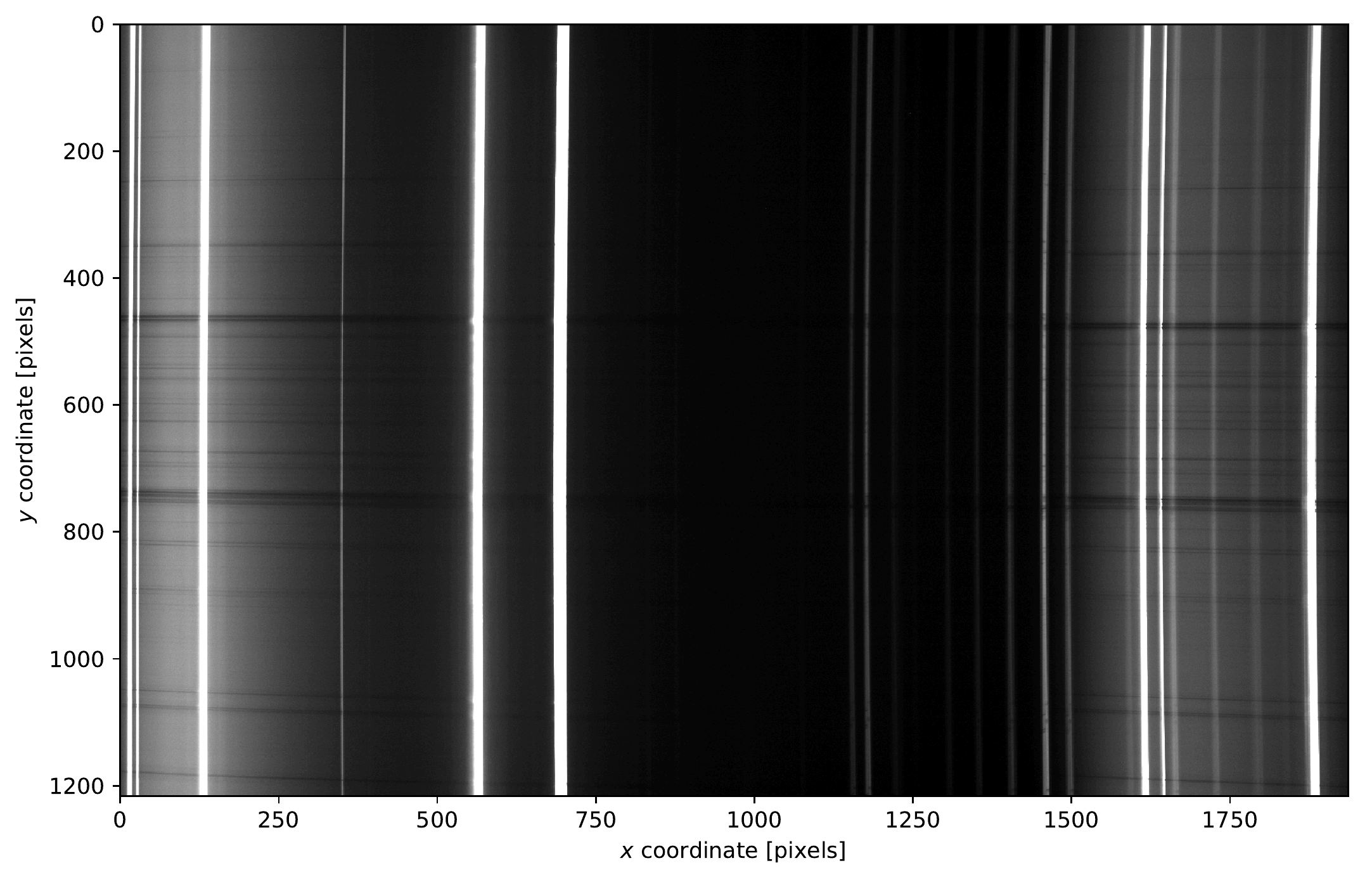}
    \caption{Image of a Hg calibration lamp spectrum taken by the long-slit grating spectrograph described in Appendix \ref{sec:appendix_spec}. The axes show the coordinate convention which we will use. Note that the pixel values in this image have been scaled to make the fainter spectral lines more visible.}
    \label{fig:ex_spec_img}
\end{figure}

Consider an image of a spectrum, such as the one in Figure \ref{fig:ex_spec_img}, which is an image of a Hg calibration lamp spectrum. This image was taken by our spectrograph described in Appendix \ref{sec:appendix_spec}. Let the spectral direction (along pixel rows) be called the $x$ direction, and let the spatial direction (along pixel columns) be called the $y$ direction. Let $W$ and $H$ be the width and height of the image respectively in pixels. Each pixel in the image can be identified by a unique coordinate pair $(x,y)$, where $x\in[0,W-1]$ and $y\in[0,H-1]$. Let us take the origin $(0,0)$ to be at the top left corner of the image. $x$ values will increase moving rightwards, and $y$ values will increase moving downwards. Wavelength $\lambda$ will increase as we move from left to right in the image.

\begin{figure}[H]
    \centering
    \includegraphics[width=0.5\textwidth]{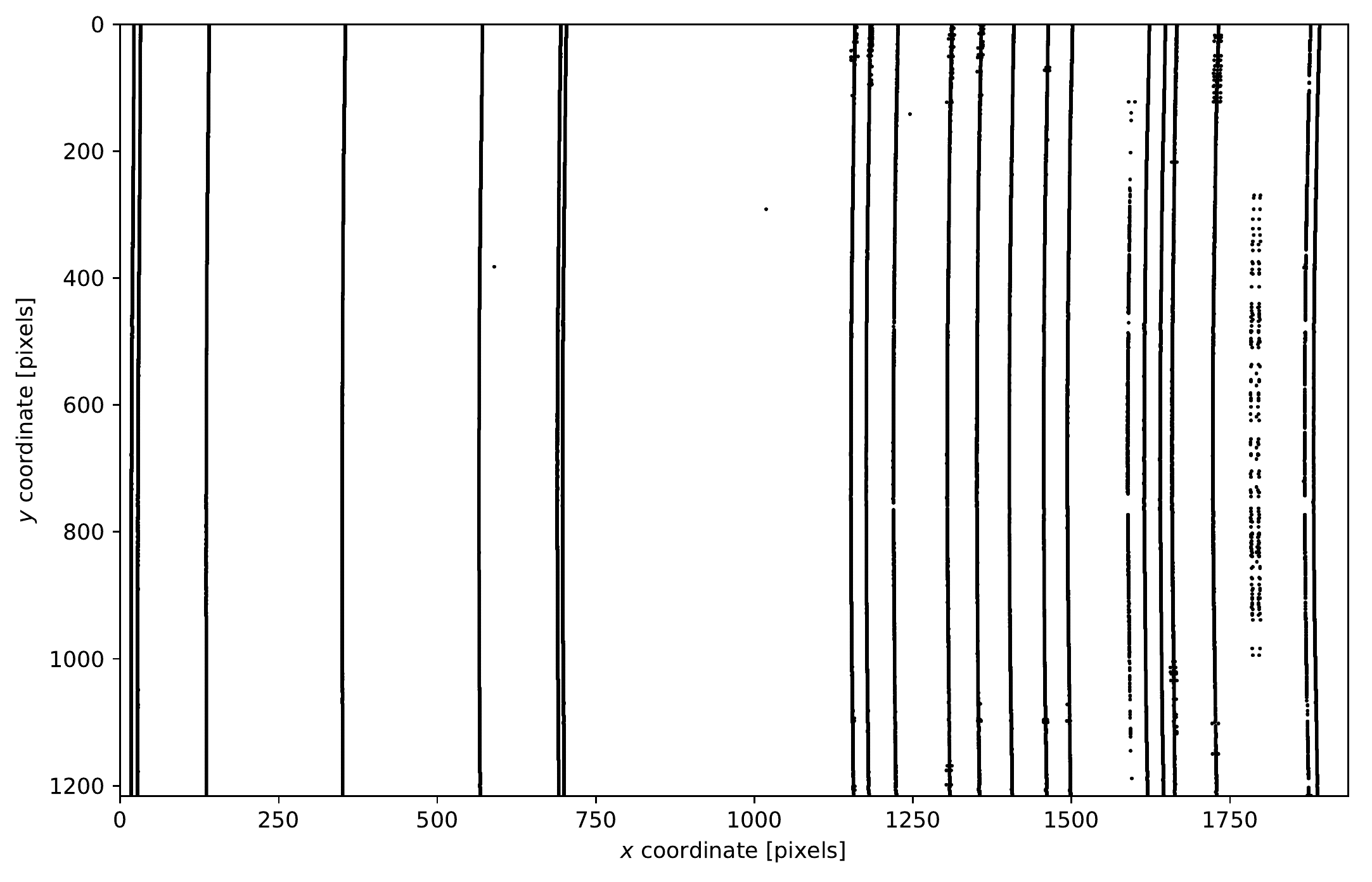}
    \caption{The set of points $S$ which corresponds to the positions of spectral peaks found in a Hg calibration lamp spectrum taken by the spectrograph described in Appendix \ref{sec:appendix_spec}.}
    \label{fig:ex_spec_img_peaks}
\end{figure}

The first step of hyperspectral imaging distortion correction is to find the set of points $S$ which correspond to the positions of spectral peaks in the image. Let $K$ be the number of spectral lines in the image which we want to use for distortion correction. For each integer row number $r\in[0,H-1]$, a slice of the image was taken at $y=r$ along the $x$ direction. The maxima along the slice were then roughly estimated based on the prominence of peaks. The positions of these maxima were then refined by fitting Gaussian functions to the slice, around the positions of the estimated maxima. For maxima which were close together, a Gaussian mixture distribution (sum of Gaussian functions) was fitted to the corresponding part of the slice. Suppose that the peaks for the slice taken at $y=r$ were found to be at $x=x_{r1}, x_{r2}, \ldots, x_{rK_{r}}$, for a total of $K_r$ peaks, where $K_r \leq K$. Then let the set of peak coordinates for this slice be $S_r = \{(x_{r1},r), (x_{r2},r), \ldots (x_{rK_{r}},r)\}$. Finally, the sets $S_r$ were combined to form $S$; that is, $S = \bigcup_{r=0}^{H-1} S_r$. Note that in order to have more visible lines, the results from several images taken by our spectrograph with different exposure times were combined together to obtain the actual set $S$. Figure \ref{fig:ex_spec_img_peaks} shows the final set of points $S$ for the Hg calibration lamp spectrum taken by our spectrograph.

\section{Setting up the Problem}\label{sec:problem}

With the set of spectral peak positions $S$ identified, the next step is to assign each point in $S$ to its corresponding spectral line. For each spectral line, we want to find the subset of $S$ corresponding to it. This is the goal of the $K$-parabolas algorithm. This section will outline the setup of the problem. Since the spectral lines are roughly parabolic in shape (as from Equation (\ref{eq:delta_beta})), it is reasonable to represent each spectral line with the equation $x = a(y-h)^2+k$, where $a$, $h$, and $k$ are some parameters. Based on how we defined the origin in the image, we can make the following three observations: (1) $a > 0$ for all parabolas; (2) $a$ values increase as we go from left to right in the image; and (3) $k$ values increase as we go from left to right in the image. These observations are constraints for any spectra.

Let us formally define the problem which we want to solve. We are given a set $S = \{(x_1,y_1), \ldots, (x_N,y_N)\}$ containing $N$ points, where $S\subset \mathbb{R}^2$. Our goal is to cluster these $N$ points into $K$ parabola-shaped disjoint clusters $C_1,\ldots,C_K$, where each point belongs to the parabola which is the closest to it. Each of the $K$ parabola-shaped clusters represents one spectral line in the image. Let us label the parabolas by the integer variable $i \in [1,K]$. The $i$th parabola can be described by the equation:
\begin{equation}
    x = a_i (y-h_i)^2 + k_i  
\end{equation}
or by the tuple $(a_i,h_i,k_i)$ representing the parabola's parameters. $C_i$ is the set representing the points assigned to the $i$th parabola (note that $S = \bigcup_{i=1}^{K} C_i$). Let the leftmost and rightmost parabolas correspond to $i=1$ and $i=K$ respectively. Based on the three observations, the following two constraints must be satisfied:
\begin{itemize}
  \setlength\itemsep{0em}
  \item $a_i < a_{i+1} \:\forall i\in[1,K-1]$
  \item $k_i < k_{i+1} \:\forall i\in[1,K-1]$
\end{itemize}

The search space for the parabola parameters can be further reduced by imposing limits for the parameters. By inspecting Figure \ref{fig:ex_spec_img} in an image editing or viewing software, an upperbound for $a_i$ can be estimated. Let this be called $a_{\textrm{max}}$. Estimates can also be made for the $k_i$ and $h_i$ values. Let the lowerbound and upperbound for $k_i$ be $k_{\textrm{min}}$ and $k_{\textrm{max}}$ respectively, and let the lowerbound and upperbound for $h_i$ be $h_{\textrm{min}}$ and $h_{\textrm{max}}$ respectively. Thus, the following three additional constraints can be imposed:
\begin{itemize}
  \setlength\itemsep{0em}
  \item $0 < a_i < a_{\textrm{max}} \:\forall i\in[1,K]$
  \item $k_{\textrm{min}} < k_i < k_{\textrm{max}} \:\forall i\in[1,K]$
  \item $h_{\textrm{min}} < h_i < h_{\textrm{max}} \:\forall i\in[1,K]$
\end{itemize}

Let the set $L = \{L_1, \ldots, L_N\}$ consist of the labels for each point in $S$. If a point $(x_j,y_j) \in S$ is determined to belong to the $i$th parabola, then we say that $L_j = i$ and $(x_j,y_j)\in C_i$. Note that $|L|=|S|$ and $L_j \in [1,K]$. Given $S$, $K$, and the five constraints described in this section, we want to find $L$ and the set of parabolas $\{(a_1,h_1,k_1), \ldots, (a_K,h_K,k_K)\}$, such that every point $(x_j,y_j)\in S$ is assigned to the nearest parabola.


\section{$K$-parabolas}\label{sec:k_parabolas}

In this section, the $K$-parabolas algorithm will be explained in detail. $K$-parabolas is an iterative algorithm which can be summarized in four simple steps as follows:
\begin{enumerate}
    \item Initialization: Select some initial values for the parabola parameters $(a_1,h_1,k_1), \ldots, (a_K,h_K,k_K)$.
    \item Cluster the Points: For every point $(x_j,y_j) \in S$, find the parabola which has the smallest orthogonal distance to it, and assign the point to that parabola-shaped cluster.
    \item Recompute the Parabola Parameters: For each parabola-shaped cluster $x = a_i (y-h_i)^2 + k_i$, use its set of assigned points $C_i$ to recompute and update its parameters $(a_i,h_i,k_i)$, so that the new parabola parameters represent the assigned points.
    \item Repeat until Convergence: Repeat steps 2 and 3 until some convergence criterion is satisfied.
\end{enumerate}


\subsection{Step 1: Initialization}\label{sec:step1_init}

In $K$-means, the first step is to initialize the centroids of the $K$ clusters, after which the centroids will be iteratively refined. Since $K$-means is an iterative algorithm, the results are sensitive to this initialization of the centroids. One simple method is to randomly choose $K$ data points in the dataset and then set those as the initial centroids of the clusters \cite{Pena1999}, but there are also other methods which produce more optimal results \cite{Steinley2007,Arthur2007}. Likewise, the first step in $K$-parabolas is to initialize the parameters for each parabola $(a_i,h_i,k_i)$. However, it is important to ensure that the constraints mentioned in Section \ref{sec:problem} are respected during this initialization. Thus, initialization methods involving random initialization are not ideal because they cannot guarantee that the constraints will be enforced.

\begin{figure}[H]
    \centering
    \begin{subfigure}{0.49\textwidth}
        \centering
        \includegraphics[width=\textwidth]{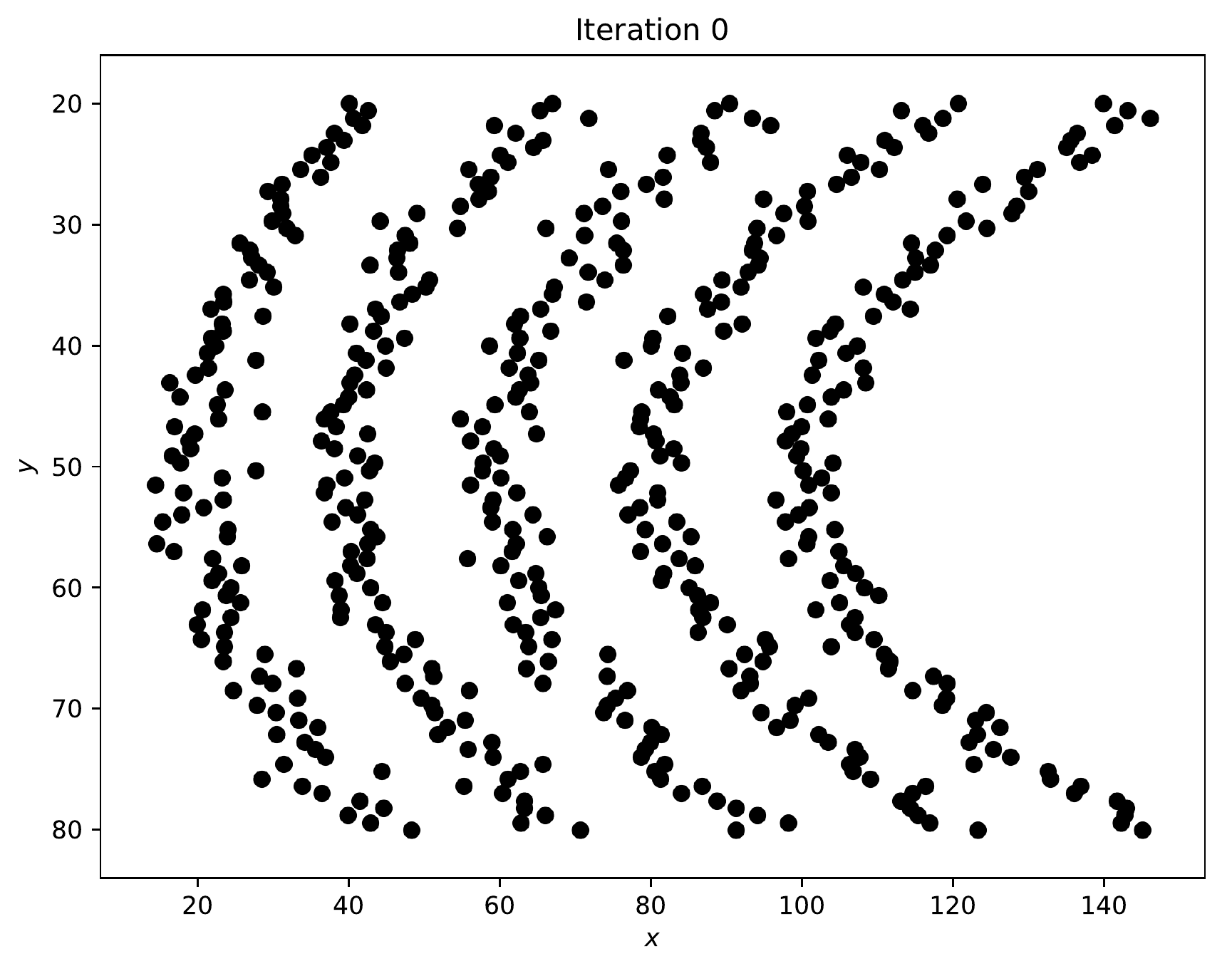}
        \caption{The given set of points $S$}
        \label{fig:step1_S}
    \end{subfigure}
    \begin{subfigure}{0.49\textwidth}
        \centering
        \includegraphics[width=\textwidth]{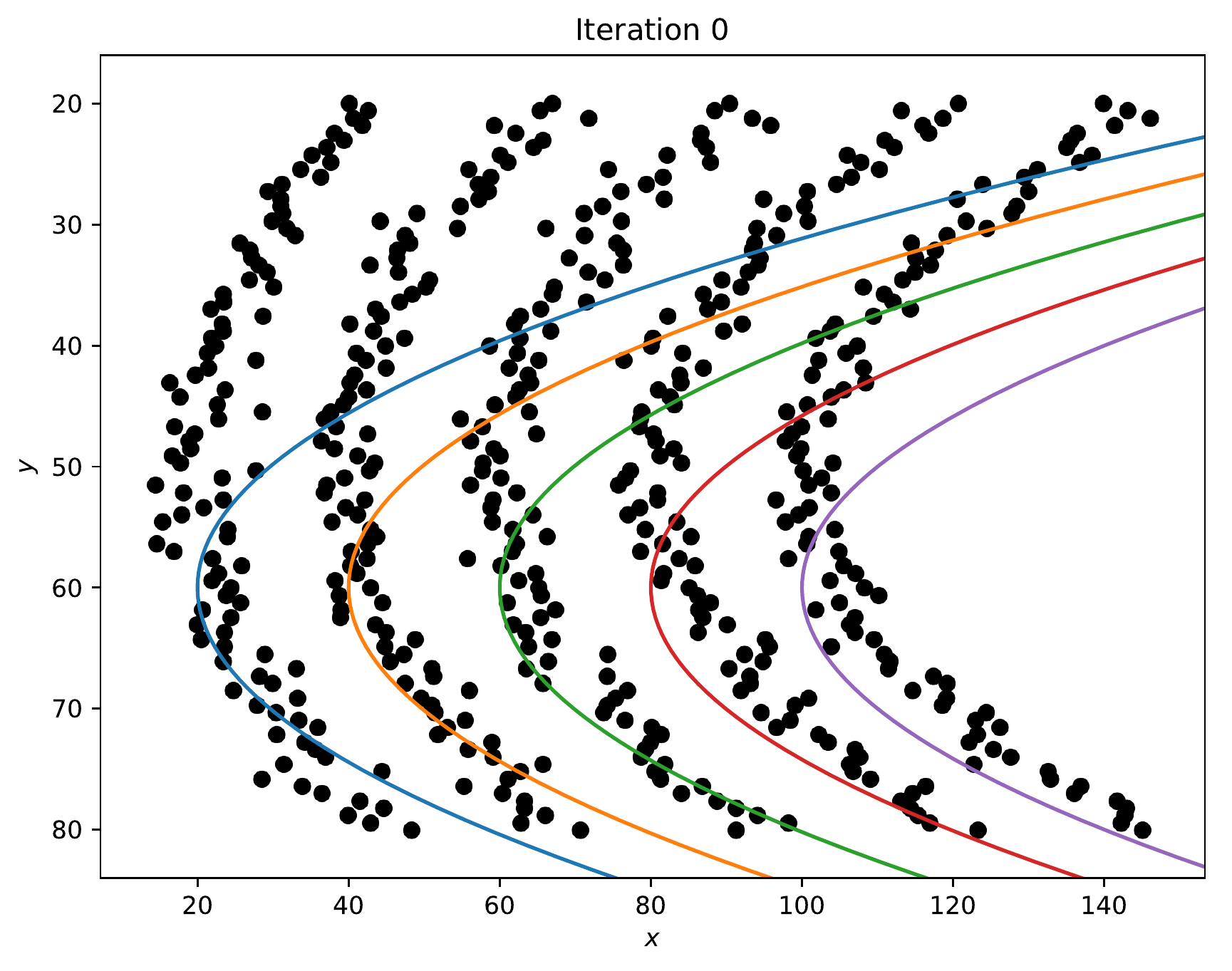}
        \caption{Initial parabolas in colour}
        \label{fig:step1_parabolas}
    \end{subfigure}
    \caption{Step 1 of $K$-parabolas: Initialize the parabola parameters.}
    \label{fig:step1}
\end{figure}

In order to ensure that the constraints are enforced, the $a_i$ values were initialized using the equation ${a_i = a_\textrm{max}(0.99)^{K-i}}$. There are also other possible initialization approaches for the $a_i$ parameters, but this ensures that $a_i < a_{i+1} \:\forall i\in[1,K-1]$. The $h_i$ and $k_i$ values, which define each parabola's vertex, were initialized based on user input. The user chooses a set of initial parabola vertices $V_\textrm{init} = \{(h_1,k_1), \ldots, (h_K,k_K)\}$. It is not required that $V_\textrm{init} \subset S$, but $V_\textrm{init}$ must satisfy $|V_\textrm{init}|=K$. One simple choice for $V_\textrm{init}$ is $V_\textrm{init} = S_r$ (see Section \ref{sec:spec_lines_id}) where $r \sim \lfloor H/2\rfloor$. This choice is likely to produce an optimal result because along the centre of the spectral image ($y\sim\lfloor H/2\rfloor$) is where all peaks are expected to be most visible, and because the actual parabola vertices are usually near the centre of the image. Should the user input $V_\textrm{init} = S_r$, then $V_\textrm{init}$ would simply be $V_\textrm{init} = \{(h_1, r), \ldots, (h_K, r)\}$. With this initialization, $K$-parabolas can more easily converge upon an optimal result. In summary, the following are provided to the $K$-parabolas algorithm by the user: $K$, $S$, $V_\textrm{init}$, $a_{\textrm{min}}$, $k_{\textrm{min}}$, $k_{\textrm{max}}$, $h_{\textrm{min}}$, and $h_{\textrm{max}}$. An example of this initialization step is shown in Figure \ref{fig:step1}, on a toy dataset of $N=500$ points in Figure \ref{fig:step1_S}. Figure \ref{fig:step1_parabolas} shows $K=5$ initial parabolas.
 
\subsection{Step 2: Cluster the Points}\label{sec:step2}

\begin{figure}[H]
    \centering
    \begin{subfigure}{0.49\textwidth}
        \centering
        \includegraphics[width=\textwidth]{Figures/test1/000000_1.pdf}
        \caption{Before}
        \label{fig:step2_1}
    \end{subfigure}
    \begin{subfigure}{0.49\textwidth}
        \centering
        \includegraphics[width=\textwidth]{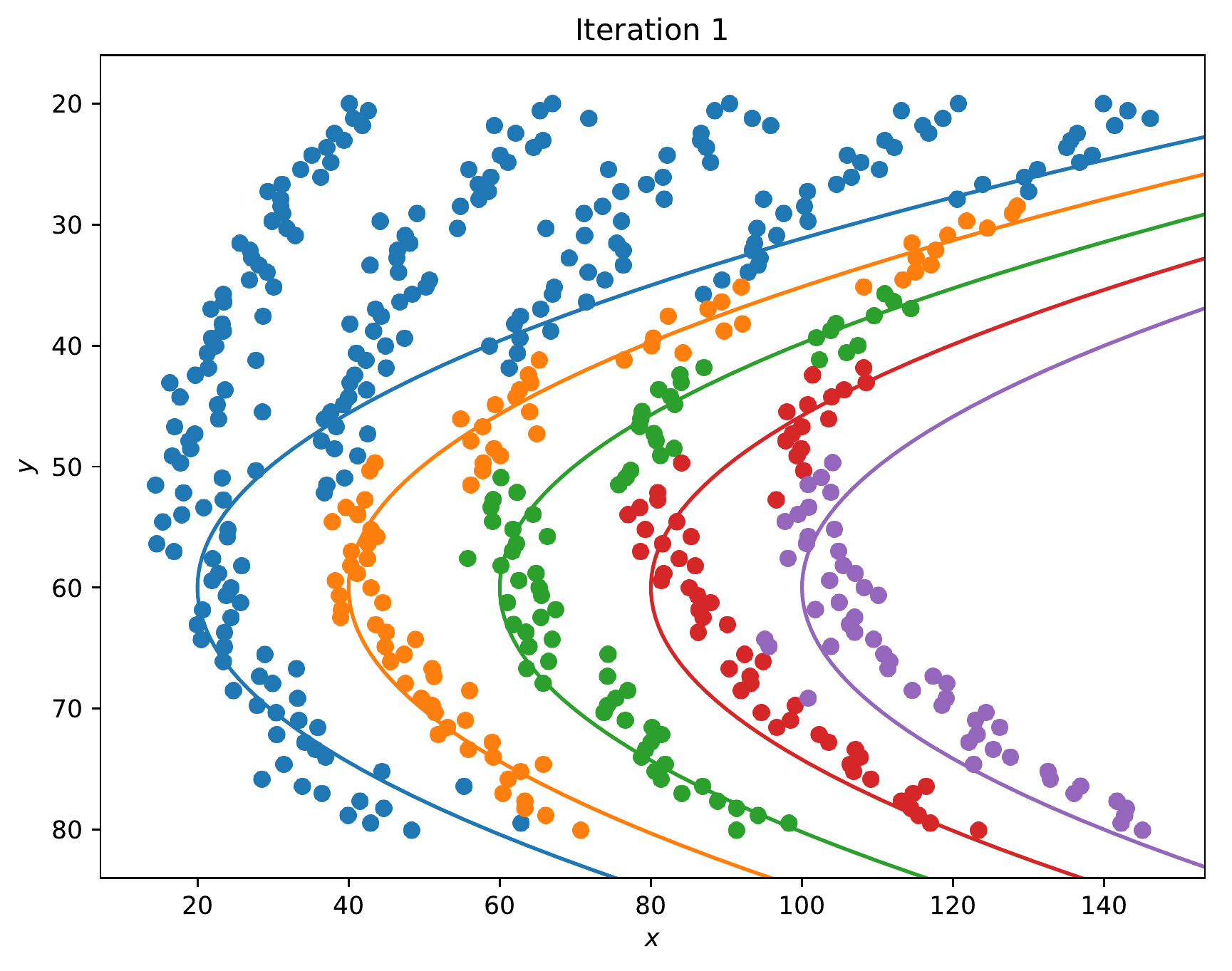}
        \caption{After}
        \label{fig:step2_2}
    \end{subfigure}
    \caption{Step 2 of $K$-parabolas: Each point $(x_j,y_j)\in S$ is assigned to the parabola closest to it, according to orthogonal distance. The points in Figure \ref{fig:step2_2} are coloured according to their assigned parabola.}
    \label{fig:step2}
\end{figure}

The next step is to cluster the points in $S$, so that each point belongs to the parabola which is the closest to it, based on orthogonal distance. For each point $(x_j,y_j) \in S$, we want to assign to it a label $L_j$, which corresponds to the parabola which has the minimum orthogonal distance to that point. Formally, for each $j \in [1,N]$, we set:
\begin{equation}
    L_j = \argmin_{i\in[1,K]}{f_i (x_j, y_j)}
\end{equation}
where $f_i (x_j, y_j)$ is the objective function, which is the orthogonal distance between point $(x_j,y_j)$ and the $i$th parabola. Then for each $i\in [1,K]$, the set $C_i$ consists of points $(x_j,y_j)$ such that $L_j = i$. An example of this step is shown in Figure \ref{fig:step2}.

\subsubsection{Calculating the Orthogonal Distance}

We want to find the minimum distance between the parabola $x=a_i(y-h_i)^2 + k_i$ and a point $(x_j,y_j) \in S$. This is the orthogonal distance between $(x_j,y_j)$ and the parabola. The distance between $(x_j,y_j)$ and any point $(x,y)$ on the parabola $x=a_i(y-h_i)^2 + k_i$ is:
\begin{equation}\label{eq:distance}
    D_i(y, x_j, y_j) = \sqrt{(y-y_j)^2 + \left([a_i(y-h_i)^2 + k_i]-x_j\right)^2}
\end{equation}
We seek a value of $y$ such that the function $D_i(y,x_j,y_j)$ is globally minimized. That is, we seek a $y^\star$ such that $\frac{\partial}{\partial y}D_i(y^\star,x_j,y_j) = 0$ and $D_i(y^\star,x_j,y_j) \leq D_i(y,x_j,y_j) \:\forall y\in\mathbb{R}$. Since the square root is a one-to-one function, $D_i(y,x_j,y_j)$ can be minimized by instead minimizing the more simple function $\Tilde{D_i}(y,x_j,y_j) \equiv [D_i(y,x_j,y_j)]^2$. For convenience, let us define the variables $b_i \equiv -2a_i h_i$ and $c_i = a_i h_i^2 + k_i$. Then $\Tilde{D_i}(y,x_j,y_j)$ can be written as:
\begin{equation}\label{eq:distance2}
    \Tilde{D_i}(y,x_j,y_j) = (y-y_j)^2 + \left(a_i y^2 + b_i y + c_i - x_j\right)^2
\end{equation}
$\Tilde{D_i}(y,x_j,y_j)$ can be minimized with respect to $y$ by solving the following equation:
\begin{equation}\label{eq:dc}
    \frac{\partial}{\partial y}\Tilde{D_i}(y,x_j,y_j)
    = 2(y-y_j) + 2(2a_i y + b_i) (a_i y^2 + b_i y + c_i - x_j) = 0 
\end{equation}
Equation (\ref{eq:dc}) is a third order polynomial equation in $y$ with three roots. Let $y^\star$ be a real root of Equation (\ref{eq:dc}) which satisfies $\Tilde{D_i}(y^\star,x_j,y_j) \leq \Tilde{D_i}(y,x_j,y_j) \:\forall y\in\mathbb{R}$. There could be at most two unique $y^\star$ which satisfy this, but either are acceptable since the corresponding values for
$\Tilde{D_i}(y^\star,x_j,y_j)$ are the same. The orthogonal distance between a point $(x_j,y_j) \in S$ and the parabola $x=a_i(y-h_i)^2 + k_i$ is given by the function:
\begin{align}\label{eq:f}
    f_i (x_j, y_j) \equiv D_i(y^\star, x_j, y_j)
\end{align}


\subsubsection{Finding the Orthogonal Distance Numerically}

Our implementation of the $K$-parabolas algorithm was done in the Python programming language. In order to find $y^\star$ numerically, a grid search method was first employed. Equation (\ref{eq:distance2}) was evaluated on a grid ranging from $y=y_\textrm{min}$ to $y=y_\textrm{max}$ inclusive, with $(y_\textrm{max}-y_\textrm{min}+1)$ points. $y_\textrm{min}$ and $y_\textrm{max}$ are selected by the user. The value of $y$ which gives the minimum value of $\Tilde{D_i}(y,x_j,y_j)$ on this grid was then used as an initial guess for numerically finding a real root of Equation (\ref{eq:dc}). Root finding was done using the Python SciPy library \cite{Virtanen2020}. With this preliminary initial guess from grid search, the root finding results were more accurate and took a shorter amount of time to obtain, since the initial guess was closer to the actual solution.

In $K$-means, computing the distances between each point and centroids of the clusters is computationally expensive in terms of time, especially for large datasets \cite{Na2010}. This is likewise the case for $K$-parabolas; computing the orthogonal distances is the bottleneck of this algorithm. In order to reduce the computation time of this step, unnecessary repeated calculations were not performed. If the $(a_i,h_i,k_i)$ values for the $i$th parabola in some iteration $\xi$ are the exact same as in iteration $\xi-1$, then there is no need to recompute the orthogonal distances from every point to the $i$th parabola in iteration $\xi$. Hence, the orthogonal distance values, parameter parameters, and labels for the immediately previous iteration were always saved in an array, and these were reused if a cluster remained the same.

\subsection{Step 3: Recompute the Parabola Parameters}\label{sec:step3}

After each point $(x_j,y_j)\in S$ was assigned a label $L_j$ (and assigned to the cluster $C_i$ if $L_j=i$), the next step is to recompute and update the parabola parameters. For each $i\in [1,K]$, a parabola $x=a_i(y-h_i)^2 + k_i$ was fitted to the points in $C_i$, using nonlinear least squares regression in SciPy \cite{Virtanen2020}, and with the constraints in Section \ref{sec:problem}. There are several ways to ensure that the constraints are respected while fitting. One method is to simultaneously fit all the parabolas while enforcing the inequality constraints for the $a_i$ and $k_i$ parameters. This was attempted using the Python LMFIT library \cite{lmfit}, but the results were too sensitive to the initial guesses. The search space could be very large for a simultaneous fit, since there are three parameters for each parabola ($a_i$, $h_i$, and $k_i$), and additional parameters which need to be defined in order to enforce the constraints. For example, to enforce the constraint $a_i < a_{i+1} \:\forall i\in[1,K-1]$, the parameters $\delta_i \equiv a_i - a_{i+1}$ were defined for each $i\in[1,K-1]$, and the constraint $\delta_i < 0 \:\forall i\in[1,K-1]$ was instead enforced. This resulted in a large search space.

\begin{figure}[H]
    \centering
    \begin{subfigure}{0.49\textwidth}
        \centering
        \includegraphics[width=\textwidth]{Figures/test1/000001_1.pdf}
        \caption{Before}
        \label{fig:step3_1}
    \end{subfigure}
    \begin{subfigure}{0.49\textwidth}
        \centering
        \includegraphics[width=\textwidth]{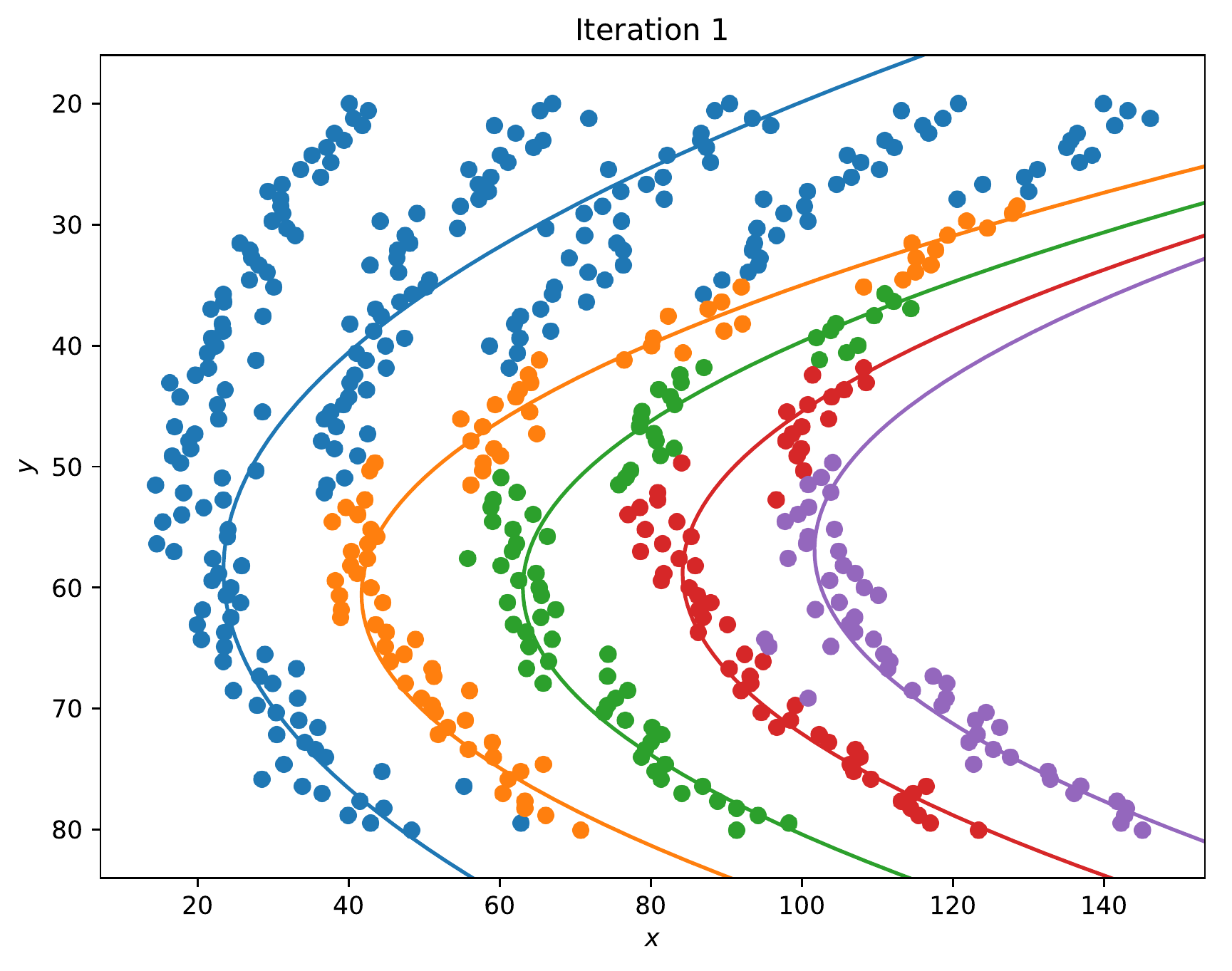}
        \caption{After}
        \label{fig:step3_2}
    \end{subfigure}
    \caption{Step 3 of $K$-parabolas: Recompute the parabola parameters based on the new assignment of points in step 2.}
    \label{fig:step3}
\end{figure}

Instead, the parabolas were fitted sequentially. Firstly, a parabola was fitted to the points in $C_1$, with the constraints $0<a_1<a_\textrm{max}$, $h_\textrm{min}<h_1<h_\textrm{max}$, and $k_\textrm{min}<k_1<k_\textrm{max}$. Next, after obtaining the new parameters $(a_1,h_1,k_1)$, a parabola was fitted to the points in $C_{2}$, with the bounds for the parameters restricted so that $a_1<a_2<a_\textrm{max}$, $h_\textrm{min}<h_2<h_\textrm{max}$, and $k_1<k_2<k_\textrm{max}$. After obtaining the new parameters $(a_2,h_2,k_2)$, a parabola was fitted to the points in $C_3$, with the bounds for the parameters restricted so that $a_2<a_3<a_\textrm{max}$, $h_\textrm{min}<h_3<h_\textrm{max}$, and $k_2<k_3<k_\textrm{max}$. This procedure was repeated for the rest of the parabolas. This is the method that we used, but other methods for fitting parabolas to the data are also possible. Figure \ref{fig:step3} shows the parabolas being refitted to the new assignment of points from step 2.

\subsection{Step 4: Repeat until Convergence}

\begin{figure}[H]
    \centering
    \includegraphics[width=0.49\textwidth]{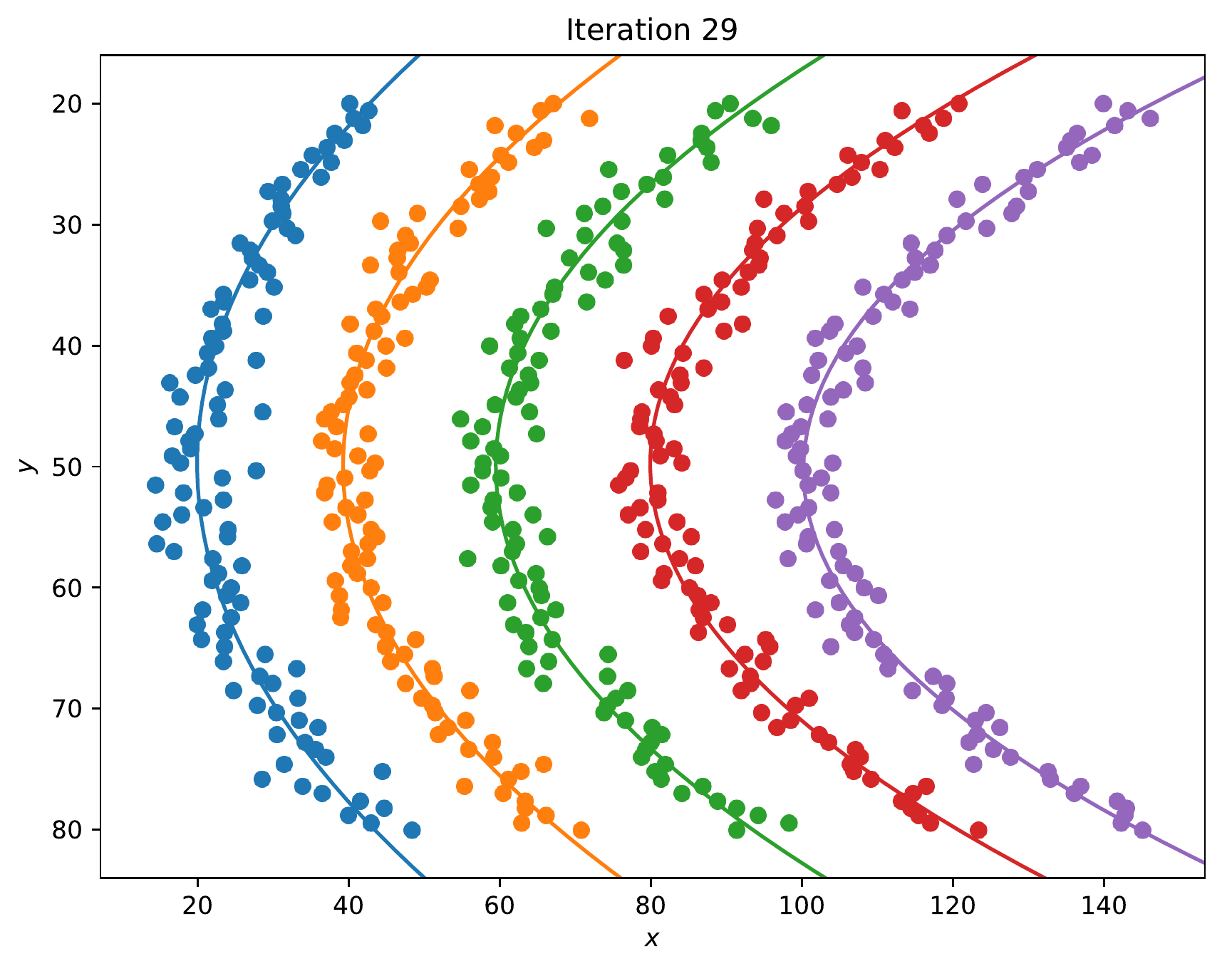}
    \caption{The final result of the $K$-parabolas algorithm applied to the toy dataset in Figure \ref{fig:step1_S}. For this example, 29 iterations were required to reach convergence.}
    \label{fig:step4}
\end{figure}

\begin{figure}[H]
    \centering
    \includegraphics[width=0.5\textwidth]{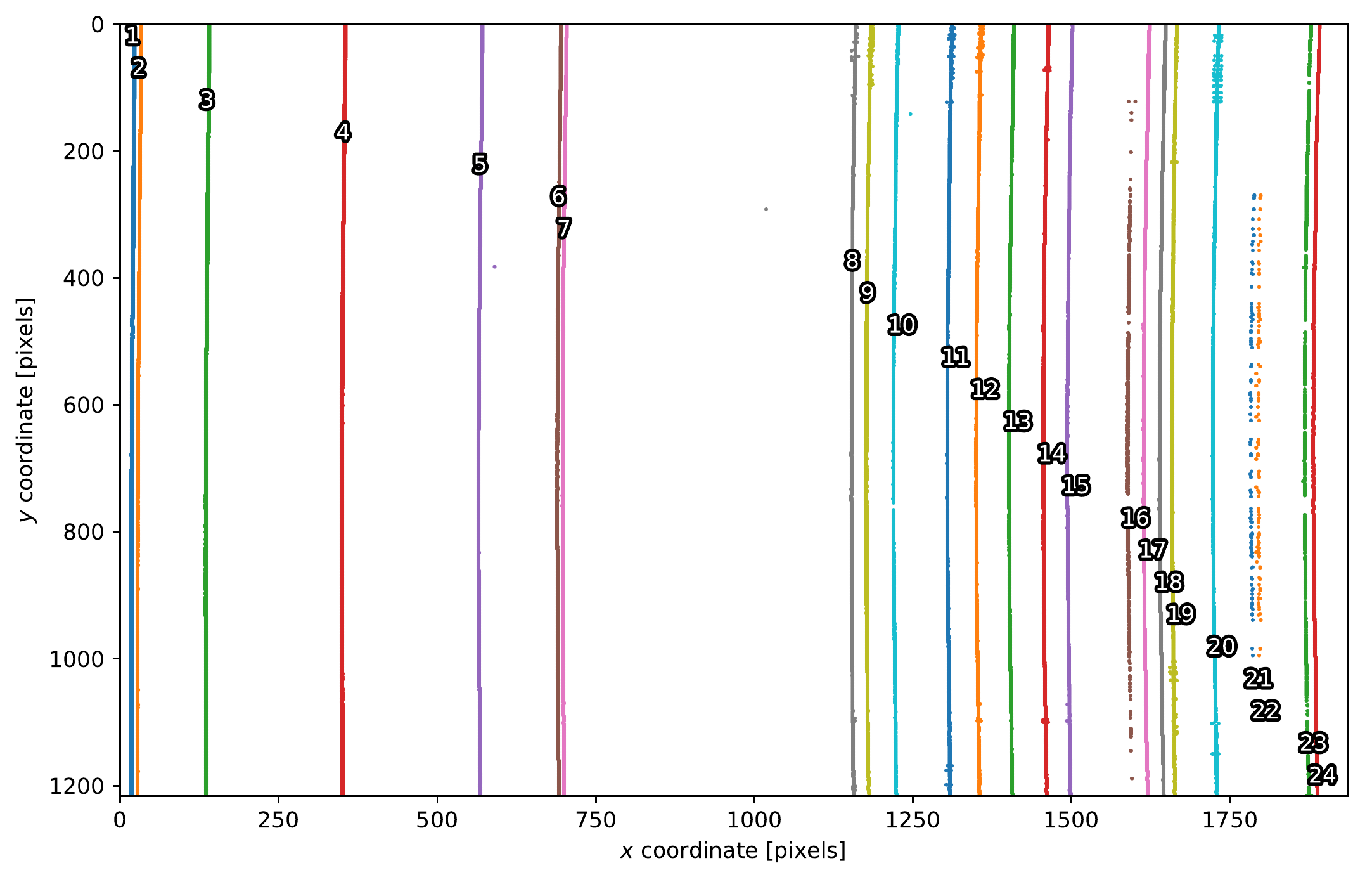}
    \caption{The final result of the $K$-parabolas algorithm for points in Figure \ref{fig:ex_spec_img_peaks}. Each cluster is numbered 1 through 24, for $K=24$ spectral lines.}
    \label{fig:ex_spec_img_clusters}
\end{figure}

Finally, steps 2 and 3 were repeated until a solution was converged upon, based on some convergence criterion. One possible convergence criterion is that the set of labels $L$ remains completely unchanged compared to the previous iteration. That is, if for every $L_j\in L$, the value $L_j$ is the same as in the previous iteration, then we could say that a solution has been converged upon. This is the criterion that we used, but like in $K$-means, there are other possible criteria. Note that when a solution has been reached, this solution is only a locally optimal solution. Hence, in $K$-means, the algorithm is often run multiple times with different initializations in order to achieve a more globally optimal solution \cite{Steinley2007}. A similar method could be employed for $K$-parabolas, and these four steps could be repeated multiple times with different initial parameters in Section \ref{sec:step1_init}. Figure \ref{fig:step4} shows the final result of the $K$-parabolas algorithm applied to the toy dataset in Figure \ref{fig:step1_S}. Figure \ref{fig:ex_spec_img_clusters} shows the final result of the $K$-parabolas algorithm applied to the set $S$ in Figure \ref{fig:ex_spec_img_peaks}. Here, $K=24$ spectral lines were identified.

\section{Refining the Parabolas and Creating a Pixel-to-Wavelength Mapping}\label{sec:refine_and_map}

\subsection{Refining the Parabolas}

\begin{figure}[H]
    \centering
    \includegraphics[width=0.45\textwidth]{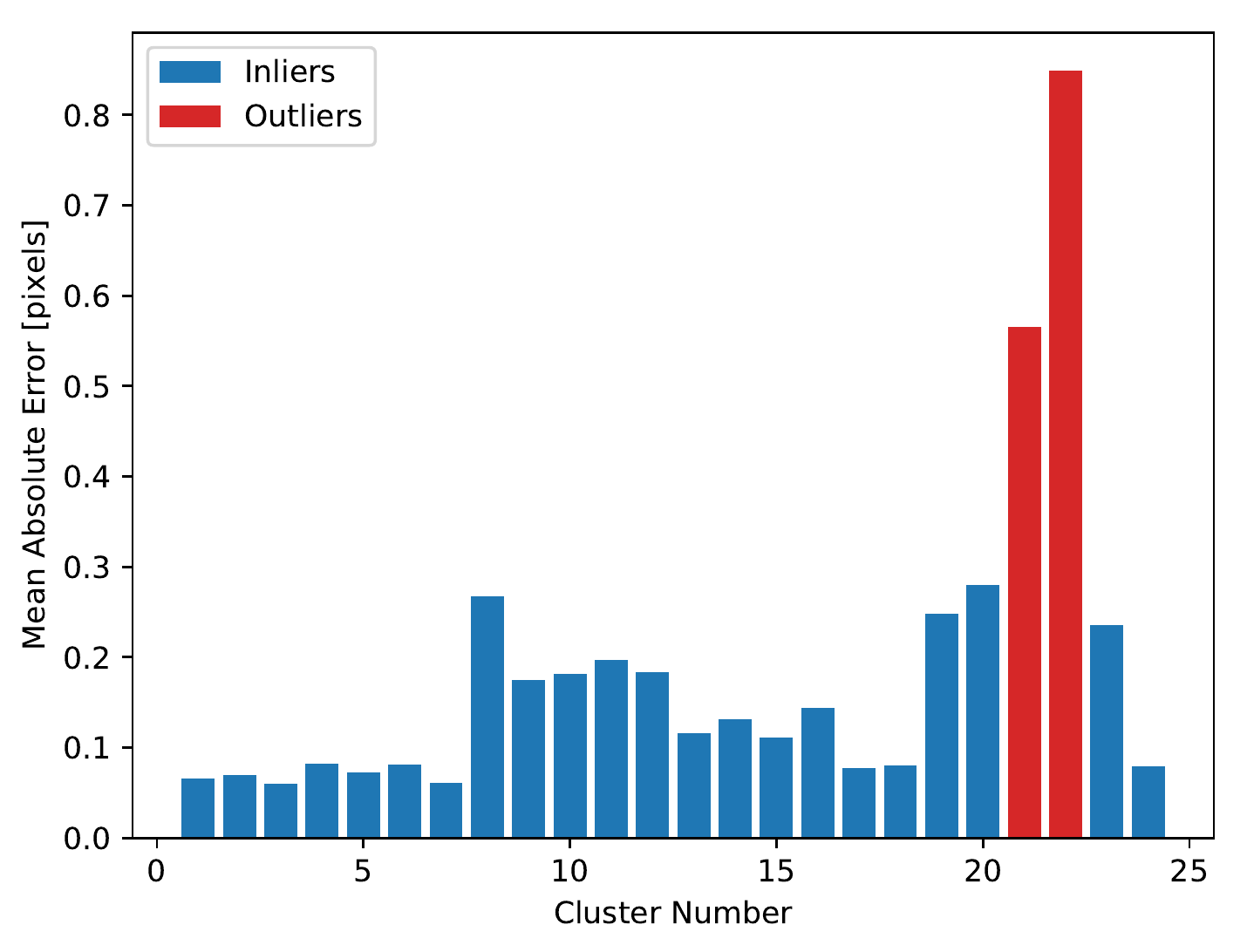}
    \caption{Mean absolute error of the refined parabolas, fit to clusters in Figure \ref{fig:ex_spec_img_clusters}.}
    \label{fig:ex_spec_img_MAE}
\end{figure}

After a solution was converged upon in $K$-parabolas, the parabolas representing each cluster were refined. The constraints mentioned in Section \ref{sec:problem} for the $a_i$ parameters could have caused $K$-parabolas to reach an inadequate local optimum. Hence, for each cluster $C_i$, a new parabola was fitted to the points in $C_i$, but instead using Huber regression. Compared to the sum of squared residuals, which is minimized in nonlinear least squares, Huber loss \cite{Huber1964}, which is minimized in Huber regression, is less sensitive to outliers in the data. However, nonlinear least squares regression, and not Huber regression, was used during step 3 of $K$-parabolas in Section \ref{sec:step3}, because the so-called ``outlier'' points at each iteration enable the parabolas to be refined and improved, towards an optimal solution. The mean absolute error (MAE) of these refined parabolas are shown in Figure \ref{fig:ex_spec_img_MAE}. Notice that there are two outliers with high MAE. These will be discussed in the following subsection.

\subsection{Removing Outlier Clusters}\label{sec:remove_outliers}

\begin{figure}[H]
    \centering
    \begin{subfigure}{0.4\textwidth}
        \centering
        \includegraphics[width=\textwidth]{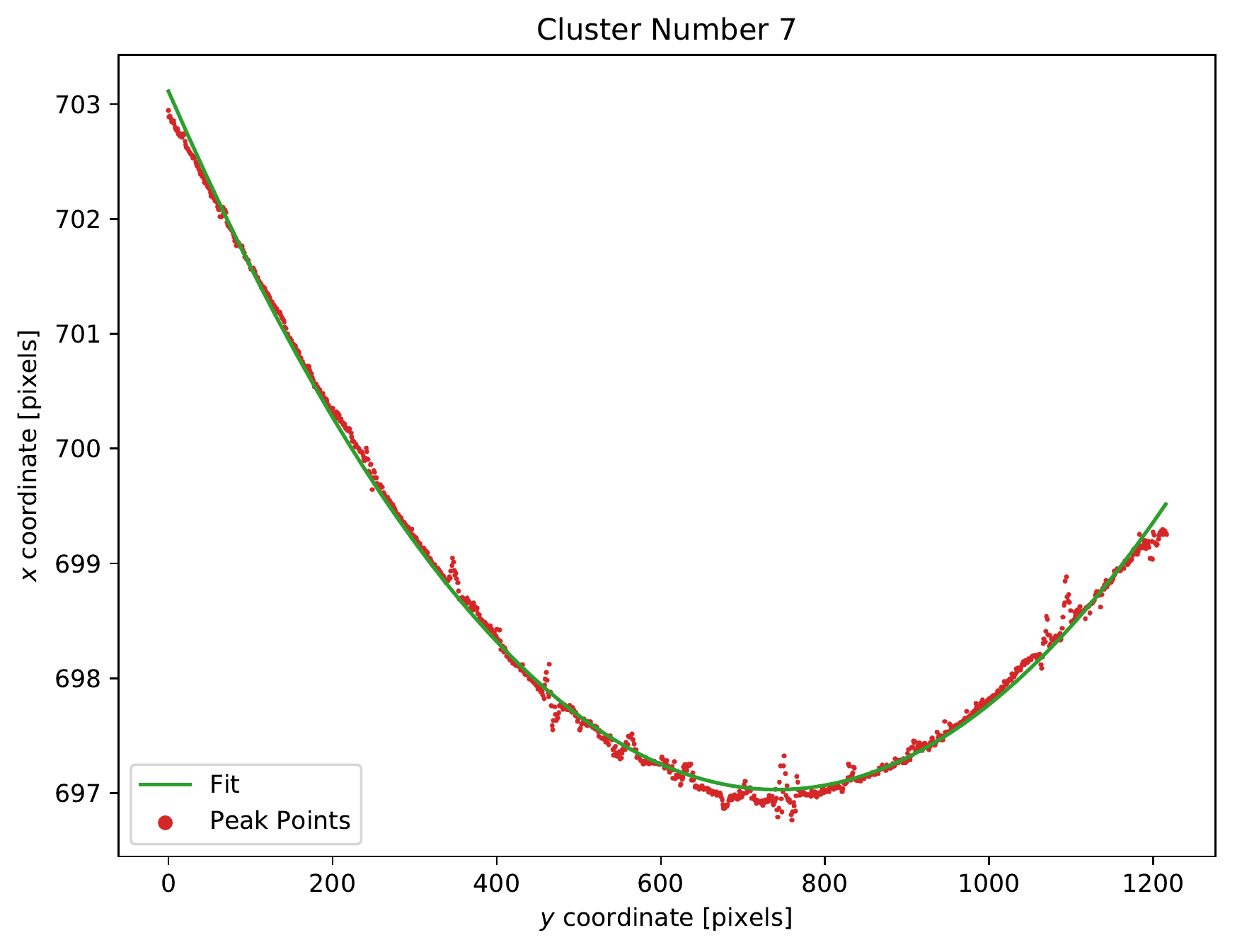}
        \caption{Cluster $C_7$; $\textrm{MAE} = 0.061$ pixels}
        \label{fig_ex_spec_img_cluster_ex_good}
    \end{subfigure}
    \begin{subfigure}{0.4\textwidth}
        \centering
        \includegraphics[width=\textwidth]{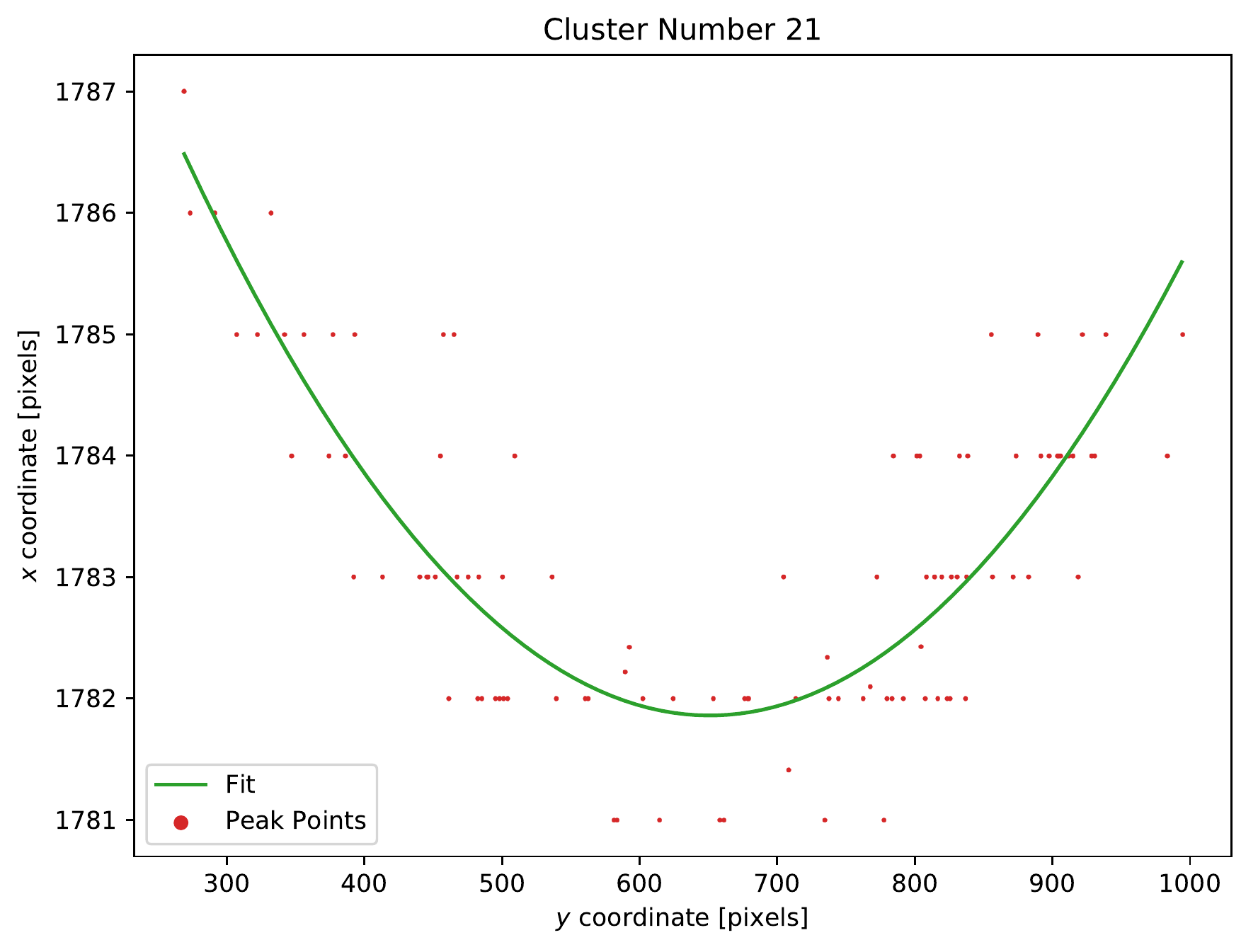}
        \caption{Cluster $C_{22}$; $\textrm{MAE} = 0.849$ pixels}
        \label{fig_ex_spec_img_cluster_ex_bad}
    \end{subfigure}
    \caption{Fitted parabolas for clusters $C_7$ and $C_{22}$ in Figure \ref{fig:ex_spec_img_clusters}. Notice that the peak positions for cluster number 22, in themselves, are poor in quality.}
    \label{fig:ex_spec_img_cluster_example}
\end{figure}

As seen in Figure \ref{fig:ex_spec_img_MAE}, there are two outliers with high MAE: $C_{21}$ and $C_{22}$. Outlier clusters with high MAE are possible in the results of $K$-parabolas, but this is not necessarily due to an inadequacy in the algorithm. Rather, this is more likely due to bad data points in the set $S$ obtained from Gaussian fitting (Section \ref{sec:spec_lines_id}), which could have been caused by bad or faint spectral lines. Figures \ref{fig_ex_spec_img_cluster_ex_good} and \ref{fig_ex_spec_img_cluster_ex_bad} respectively show the points in $C_7$ and $C_{22}$, and the parabolas representing them. The MAE for clusters 7 and 22 are 0.061 pixels and 0.849 pixels respectively. One can see that the points in cluster number 22, in themselves, are poor in quality. Cluster $C_{22}$ should be considered an outlier, and the parameters of its fitted parabola should not be used when determining a pixel-to-wavelength mapping. Looking more closely at Figures \ref{fig:ex_spec_img_peaks} and \ref{fig:ex_spec_img_clusters}, one can see that clusters $C_{21}$ and $C_{22}$ have a smaller number of points compared to the other clusters, and more gaps. The Gaussian fitting process was not as successful for these two clusters. Hence, in order to determine if a cluster is an outlier due to poor quality points, and not due to inadequacies of the $K$-parabolas algorithm, plots of the parabola fits, such as the ones in Figure \ref{fig:ex_spec_img_cluster_example}, were visually inspected. From this inspection, an appropriate threshold for the MAE of the clusters was then selected, and clusters with MAE values above this threshold were then identified as outliers, as shown in Figure \ref{fig:ex_spec_img_MAE}.

\subsection{Parabola Parameter Fits and Pixel-to-Wavelength Mapping}\label{sec:parameter_fits}

\begin{figure}[H]
    \centering
    \begin{subfigure}{0.49\textwidth}
        \centering
        \includegraphics[width=\textwidth]{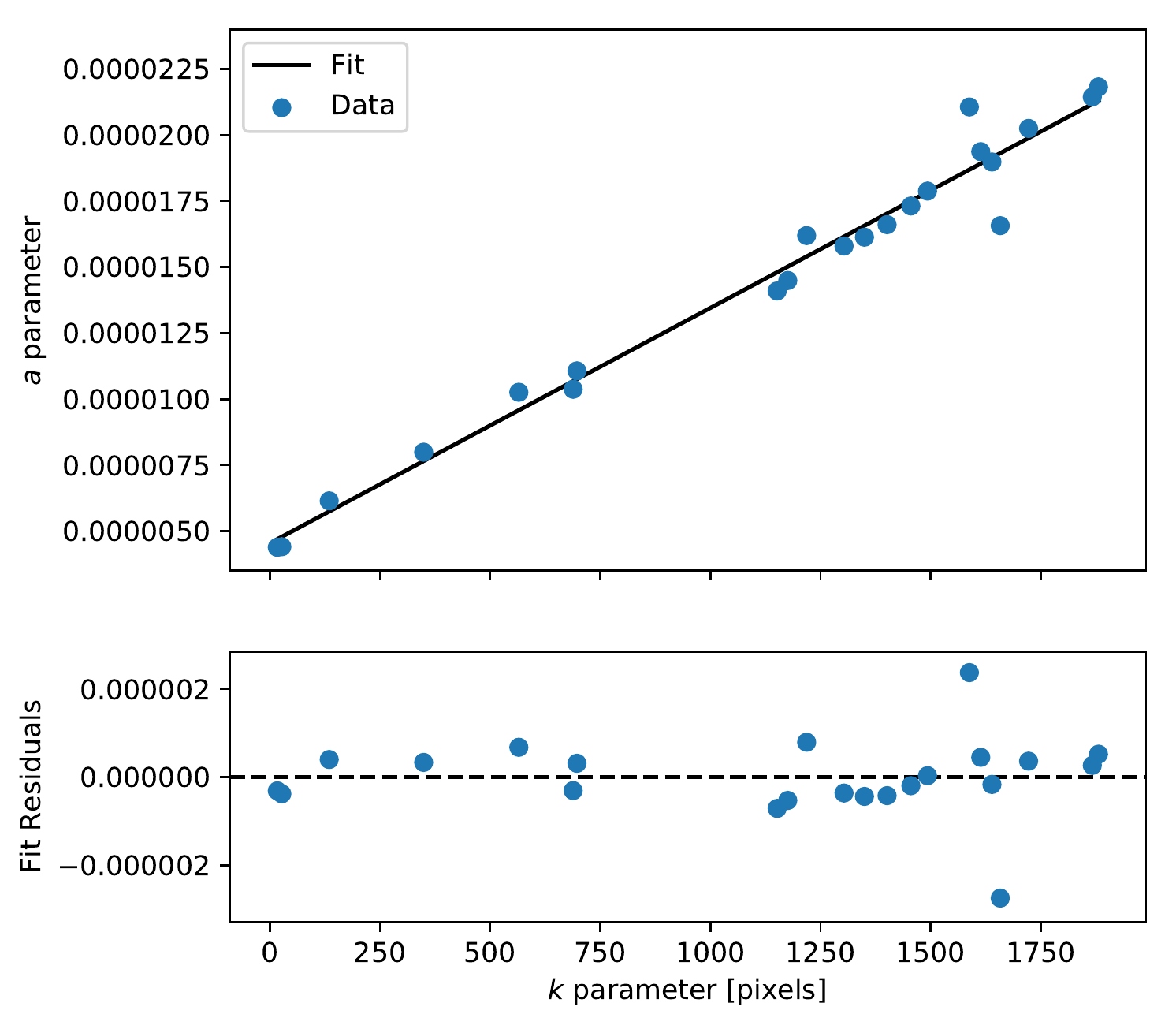}
        \caption{Linear fit of $a$ as a function of $k$}
        \label{fig:ex_spec_img_a_VS_k}
    \end{subfigure}
    \begin{subfigure}{0.49\textwidth}
        \centering
        \includegraphics[width=0.91\textwidth]{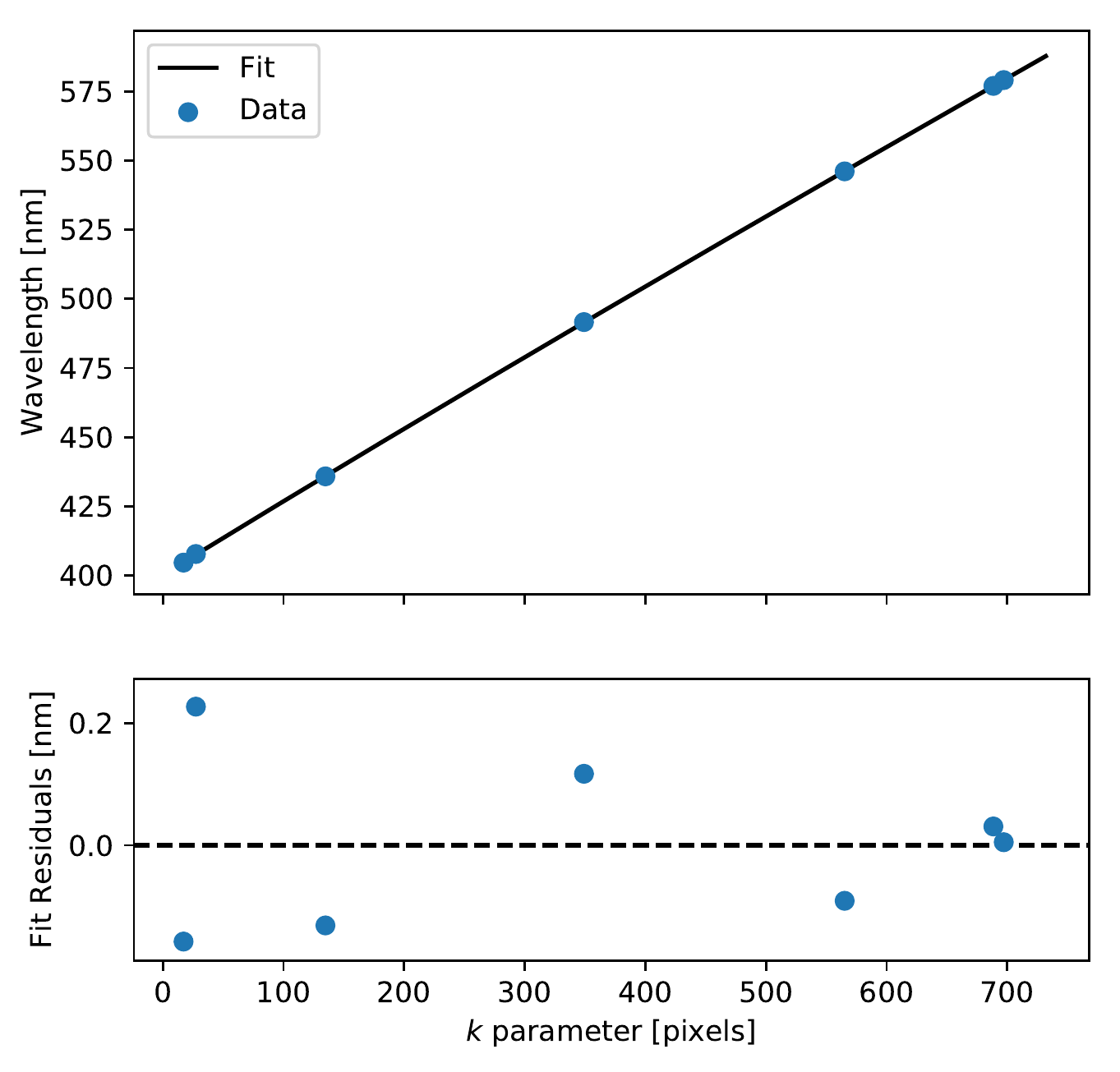}
        \caption{Cubic fit of wavelength $\lambda$ as a function of $k$}
    \label{fig:ex_spec_img_lambda_VS_k}
    \end{subfigure}
    \caption{Linear fit of $a$ as a function of $k$ for all clusters except for the identified outliers ($C_{21}$ and $C_{22}$), and cubic fit of wavelength $\lambda$ as a function of $k$ for clusters $C_{1},\ldots,C_{7}$, which correspond to first order Hg spectral lines with known wavelength values.}
    \label{fig:ex_spec_img_params_fit}
\end{figure}

At this step, for each spectral line, we now have the equation of the parabola which represents it. With this valuable information, a user can correct for hyperspectral imaging distortion using a method they choose. For example, one could apply projective transformations \cite{EsmondeWhite2011}, or could create a pixel-to-wavelength mapping, which is what we will do. A pixel-to-wavelength mapping can then be used to create a hyperspectral data cube for analyses. To do this, firstly the $a$ parameters were fitted as a linear fucntion of the $k$ parameters, the $h$ parameters were fitted as a bijective cubic function of the $k$ parameters, and wavelength $\lambda$ was fitted as a bijective cubic function of the $k$ parameters for some selected first order spectral lines with known wavelength values. Clusters $C_{1},\ldots,C_{7}$ in Figure \ref{fig:ex_spec_img_clusters} correspond to first order Hg lines, while the other lines correspond to second order Hg lines. Figure \ref{fig:ex_spec_img_a_VS_k} shows $a$ fitted as a linear function of $k$ for all clusters except for the outliers, and Figure \ref{fig:ex_spec_img_lambda_VS_k} shows $\lambda$ fitted as a cubic function of $k$ for the 7 first order Hg lines. With the bijective functions $a(k)$, $h(k)$, and $\lambda(k)$, a pixel-to-wavelength mapping can be determined. Suppose that some pixel in the image has coordinate $(x_\textrm{p},y_\textrm{p})$ where $x_\textrm{p}\in[0,W-1]$ and $y_\textrm{p}\in[0,H-1]$. There is only one parabola which $(x_\textrm{p},y_\textrm{p})$ can lie on, and the parameters of this parabola can be found by solving for the value of $k$ which satisfies the equation $x_\textrm{p}= a(k) \times (y_\textrm{p}-h(k))^2+k$. Let the value of $k$ which satisfies this equation be called $k^\star$. Then the parabola which corresponds to $(x_\textrm{p},y_\textrm{p})$ has parameters $(a(k^\star), h(k^\star), k^\star)$ and has wavelength $\lambda(k^\star)$. The amount of smile corresponding to $(x_\textrm{p},y_\textrm{p})$ is $a(k^\star) \times (y_\textrm{p}-h(k^\star))^2$ pixels.

\subsection{Results for Hg Lamp Spectrum}

\begin{figure}[H]
    \centering
    \begin{subfigure}{0.49\textwidth}
        \centering
        \includegraphics[width=\textwidth]{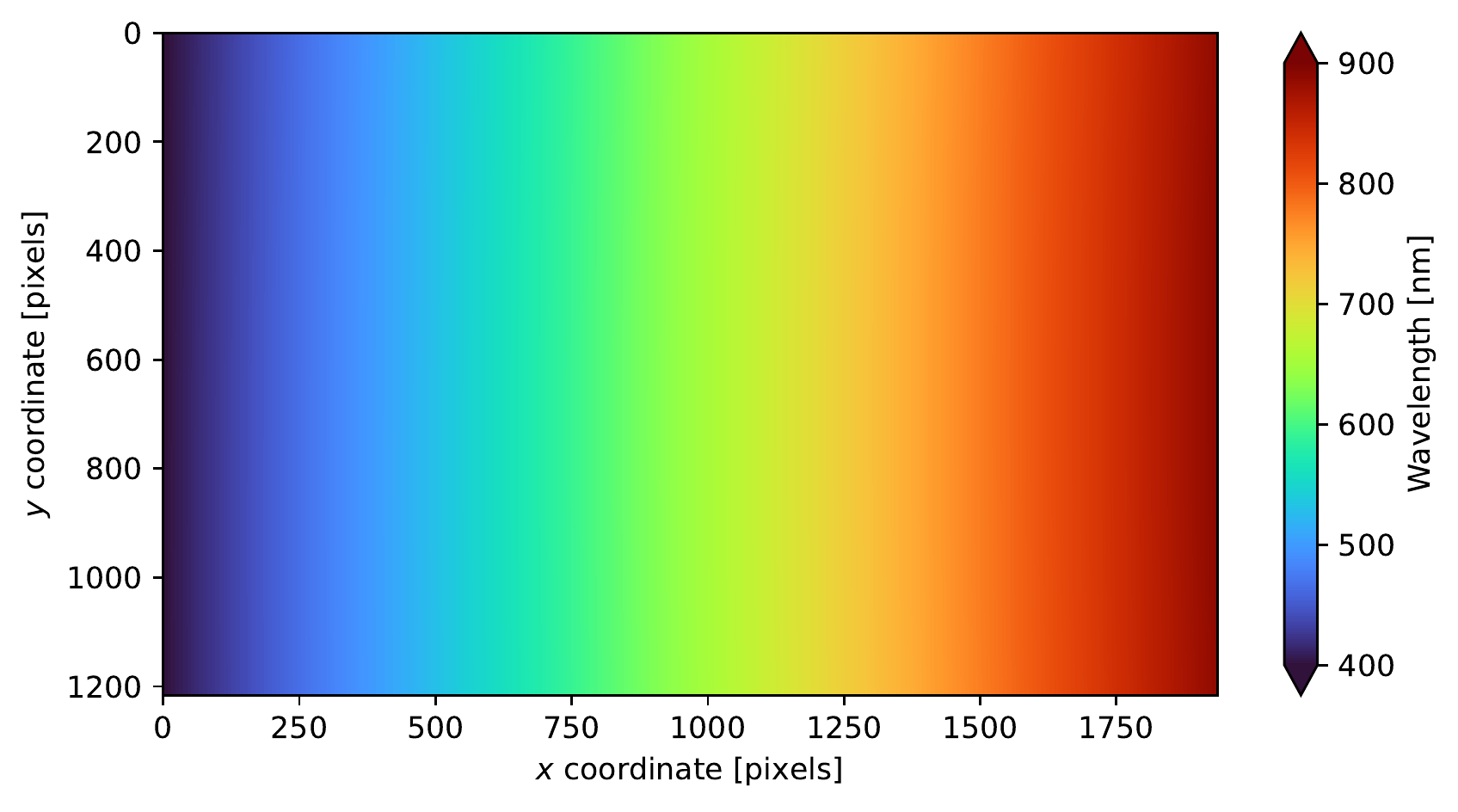}
        \caption{Pixel-to-wavelength mapping}
        \label{fig:ex_spec_img_map}
    \end{subfigure}
    \begin{subfigure}{0.49\textwidth}
        \centering
        \includegraphics[width=\textwidth]{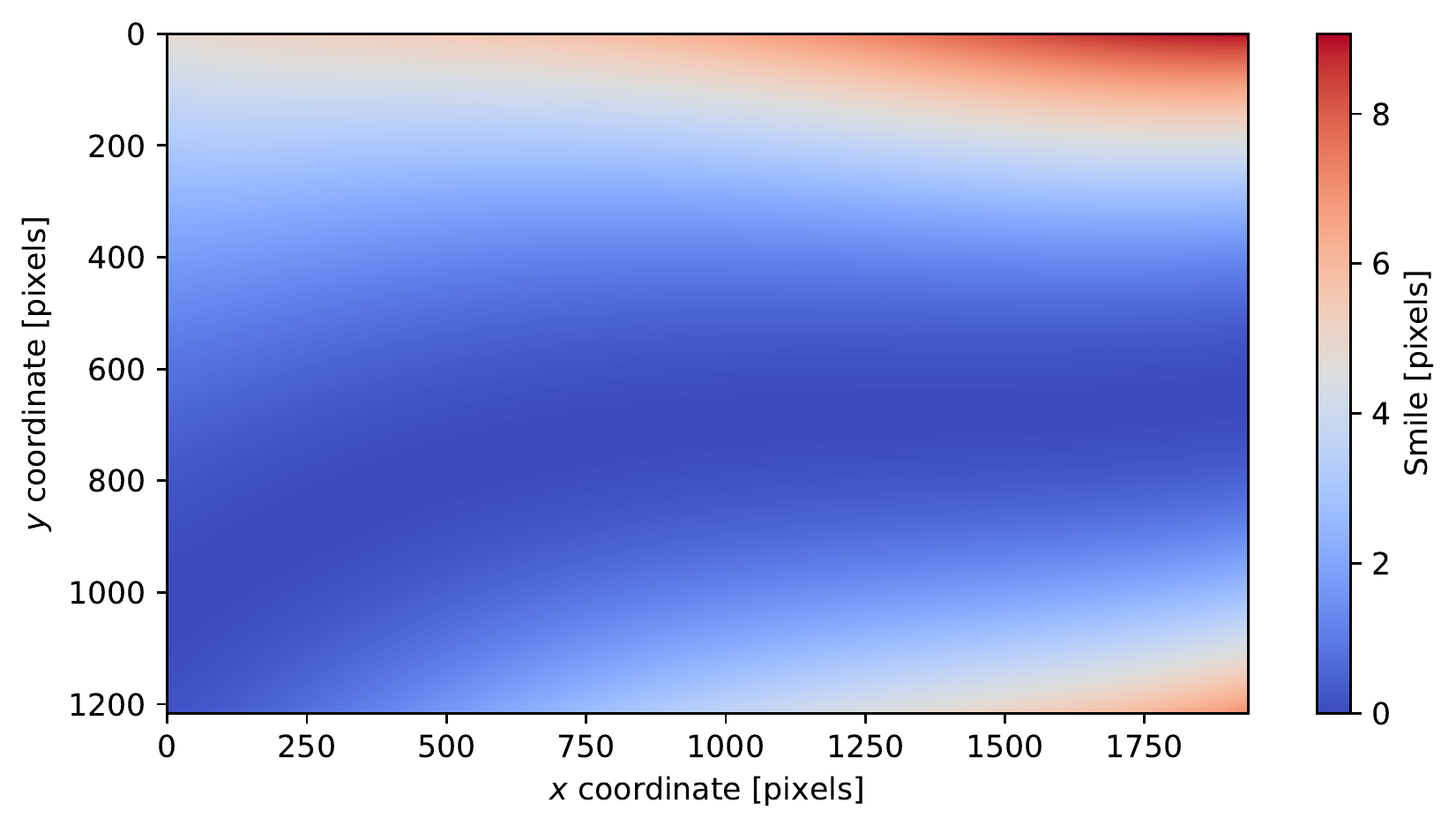}
        \caption{Amount of smile}
        \label{fig:ex_spec_img_smile}
    \end{subfigure}
    \caption{Pixel-to-wavelength mapping and amount of smile for the Hg spectrum image in Figure \ref{fig:ex_spec_img}.}
    \label{fig:ex_spec_img_result}
\end{figure}

Figure \ref{fig:ex_spec_img_result} shows the final results for the Hg lamp calibration image in Figure \ref{fig:ex_spec_img} taken by our spectrograph. Figure \ref{fig:ex_spec_img_map} shows the pixel-to-wavelength mapping, and Figure \ref{fig:ex_spec_img_smile} shows the amount of smile for each pixel. From Figure \ref{fig:ex_spec_img_smile}, we can see that smile is larger at the top and bottom edges of the image, where $|\gamma|$ is larger, and this is consistent with Equation (\ref{eq:delta_beta}). In addition, from Figure \ref{fig:ex_spec_img_smile}, we can observe that there is some tilt in the grating or image detector since the $h$ parameters increase with $y$. The final MAE of all the refined parabolas which are inliers is 0.135 pixels. As seen in Figure \ref{fig:ex_spec_img_MAE}, each inlier spectral line can be modelled as a parabola with a MAE of less than 0.3 pixels.

\newpage
\section{Extension to a DMD-Based Multi-Object Spectrograph}\label{sec:DMD-MOS}

\subsection{DMD-based Multi-Object Spectrograph Overview}

The methods described so far in Sections \ref{sec:k_parabolas} and \ref{sec:refine_and_map} have been verified on real-world data from a long-slit grating spectrograph. Let us now extend the methods to measure hyperspectral imaging distortion in a DMD-based MOS. A DMD is a microelectromechanical systems component which consists of an array of tiny mirrors, called micromirrors, that can be individually programmed to tilt in two configurations. The DMD has been used in various commercial and industrial applications\cite{Dudley2003}, and is a proven and reliable technology\cite{Travinsky2017} which has decreased in cost over the years. For these reasons, there have been several studies which have proposed the use of DMDs as the programmable slit mask of MOS systems \cite{Barkhouser2016,Meyer2004,Zamkotsian2014,Robberto2009}. At the Dunlap Institute, we are currently developing a DMD-based MOS (DMD-MOS) which will be placed on a \mbox{0.5 m} telescope as an exploratory study for future DMD-based MOS systems. At the time of writing, the design of our DMD-MOS is complete, and the components are currently being manufactured by vendors. The DMD in our DMD-MOS is a Texas Instruments DLP7000 DMD, which is an array of $H_\textrm{DMD}=1024$ by $W_\textrm{DMD} = 768$ micromirrors, each with a pitch of 13.7 \textmu m. The micromirrors can be individually programmed to be in an ON or OFF configuration, at $\pm12^\circ$. The DMD is situated at a focal plane of the telescope, and the micromirrors in the ON configuration reflect incoming light in the direction of the dispersive element, acting as slits in a conventional spectrograph. For our DMD-MOS, the dispersive element is a VPH grism. The wavelength range of our DMD-MOS is 400 nm to 700 nm. Another paper published in these proceedings discusses the optical design of our DMD-MOS \cite{Chen2022}. 

In this section, the $K$-parabolas algorithm (Section \ref{sec:k_parabolas}) and pixel-to-wavelength mapping construction procedure (Section \ref{sec:refine_and_map}) will be demonstrated on simulated data from an older design of our DMD-MOS, due to time constraints at the time of writing. The main differences between our old design and current design is that the current design has improved optics, an improved VPH grism design, and a smaller detector. Nevertheless, data from the old design is comparable to the current design, and if a method for distortion measurement works for the old design, it should also work for the new design. More information about our old design can be found in Appendix \ref{sec:appendix_DMD-MOS}. 


\subsection{Micromirror Configurations for DMD-MOS Wavelength Calibration}

The calibration of a DMD-based MOS is complex because the slit can be programmed to be in many different positions. Compared to the traditional long-slit grating spectrograph, a DMD-based MOS has two spatial directions due to the programmable slit. Depending on where the slit is, the resulting smile and keystone distortion of the spectrum will be different. In addition, as in other MOS systems, overlapping spectra from different objects must be accounted for. To calibrate our DMD-MOS, the micromirrors were programmed to have a marked slit configuration, instead of a long-slit configuration. A marked slit has periodic holes in order to separate the spatial fields, which would allow for keystone measurement. An example of this is discussed in Ref. \citenum{Hong17}, in which a ``field identifier'' was used to create a marked slit configuration for a long-slit grating spectrograph. For the DMD-MOS, a marked slit configuration was achieved by programming micromirrors to be in a spatially periodic ON configuration.

\begin{figure}[H]
    \centering
    \begin{subfigure}{0.49\textwidth}
        \centering
        \includegraphics[width=0.5\textwidth]{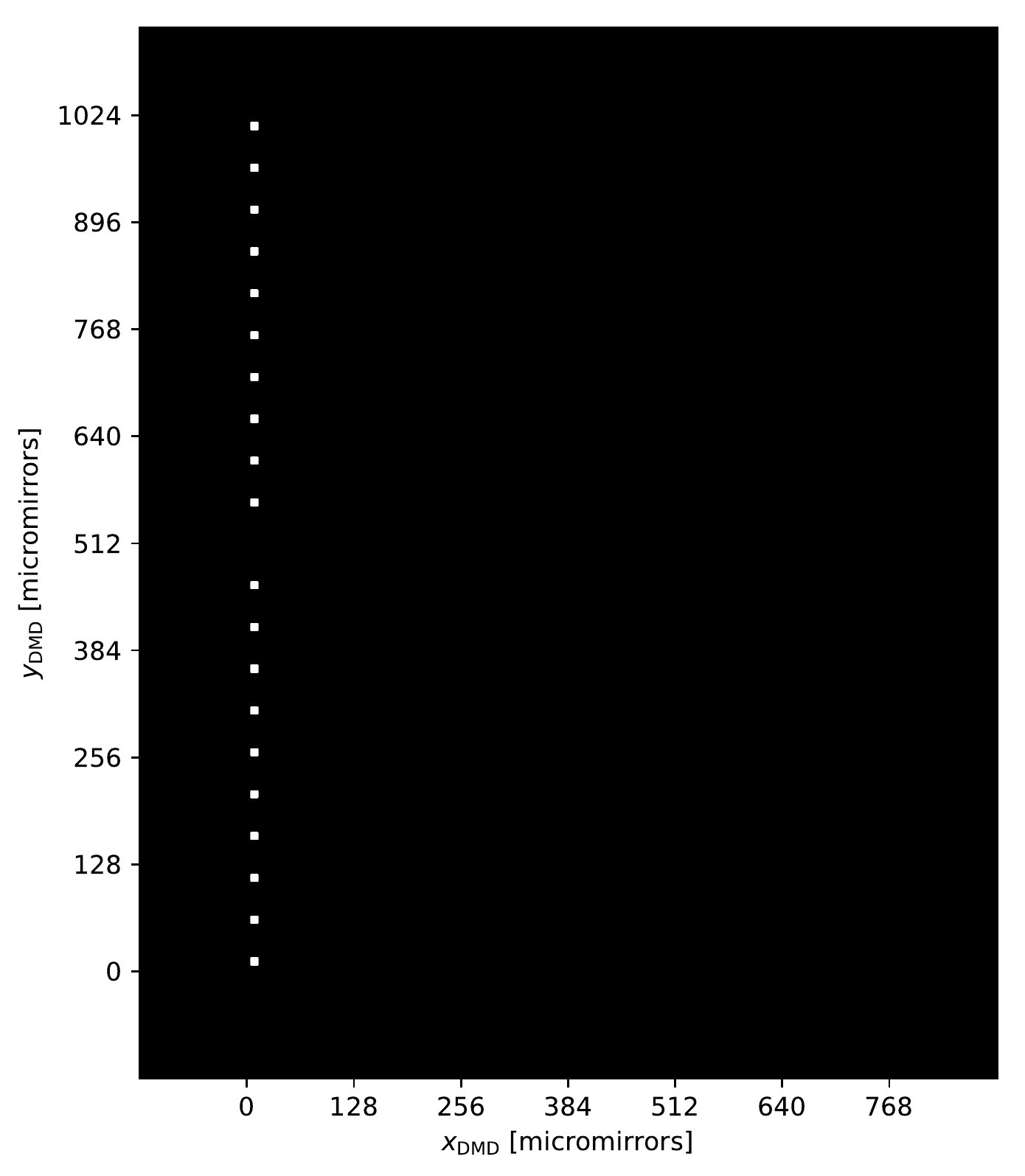}
        \caption{DMD with ON micromirrors along $x_\textrm{DMD}=9$}
        \label{fig:DMD-MOS_Footprint_DMD125}
    \end{subfigure}
    \begin{subfigure}{0.49\textwidth}
        \centering
        \includegraphics[width=0.9\textwidth]{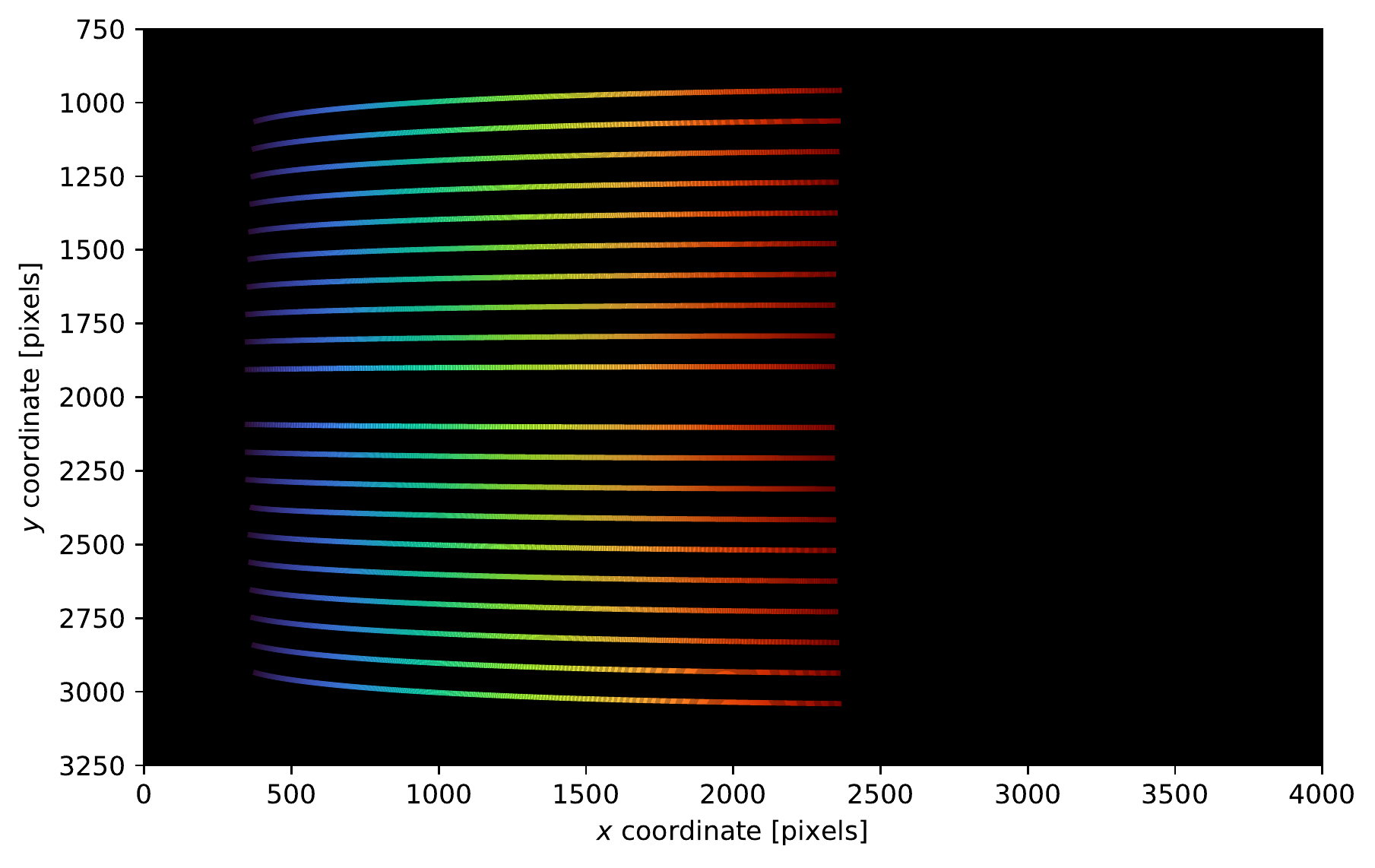}
        \caption{Corresponding spectra on detector for $x_\textrm{DMD}=9$}
        \label{fig:DMD-MOS_Footprint_Det125}
    \end{subfigure}
    \begin{subfigure}{0.49\textwidth}
        \centering
        \includegraphics[width=0.5\textwidth]{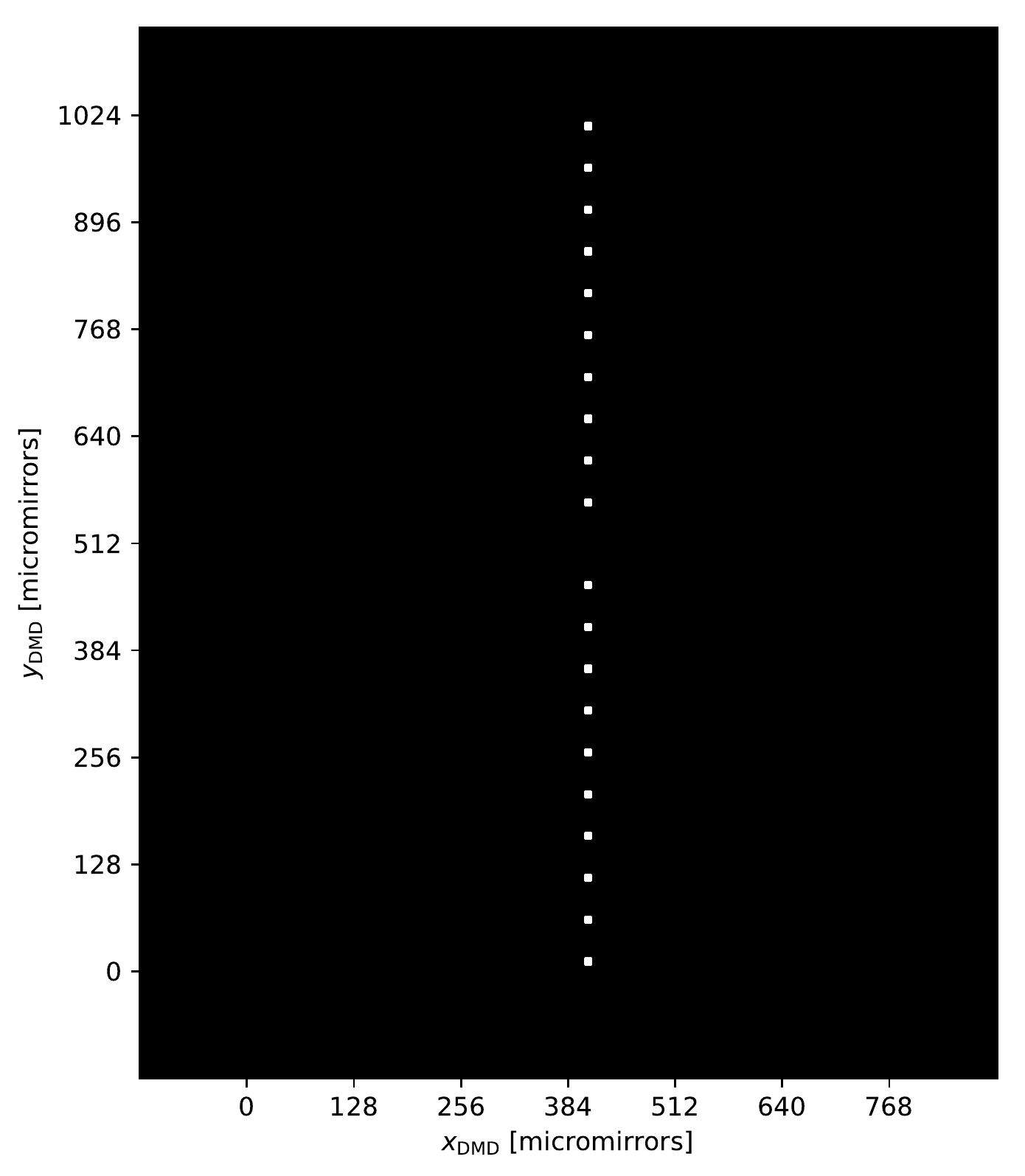}
        \caption{DMD with ON micromirrors along $x_\textrm{DMD}=408$}
        \label{fig:DMD-MOS_Footprint_DMD524}
    \end{subfigure}
    \begin{subfigure}{0.49\textwidth}
        \centering
        \includegraphics[width=0.9\textwidth]{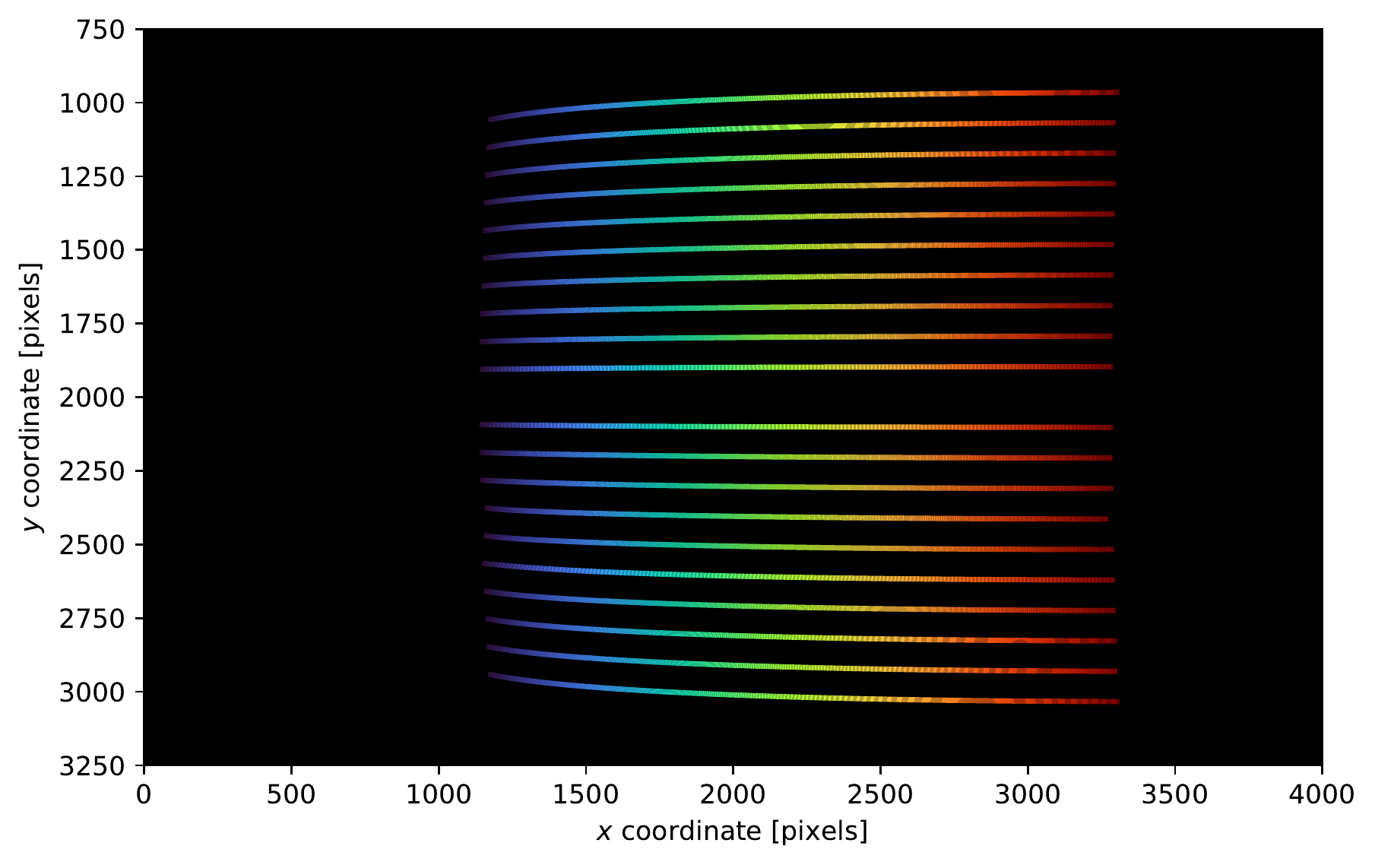}
        \caption{Corresponding spectra on detector for $x_\textrm{DMD}=408$}
        \label{fig:DMD-MOS_Footprint_Det524}
    \end{subfigure}
    \begin{subfigure}{0.49\textwidth}
        \centering
        \includegraphics[width=0.5\textwidth]{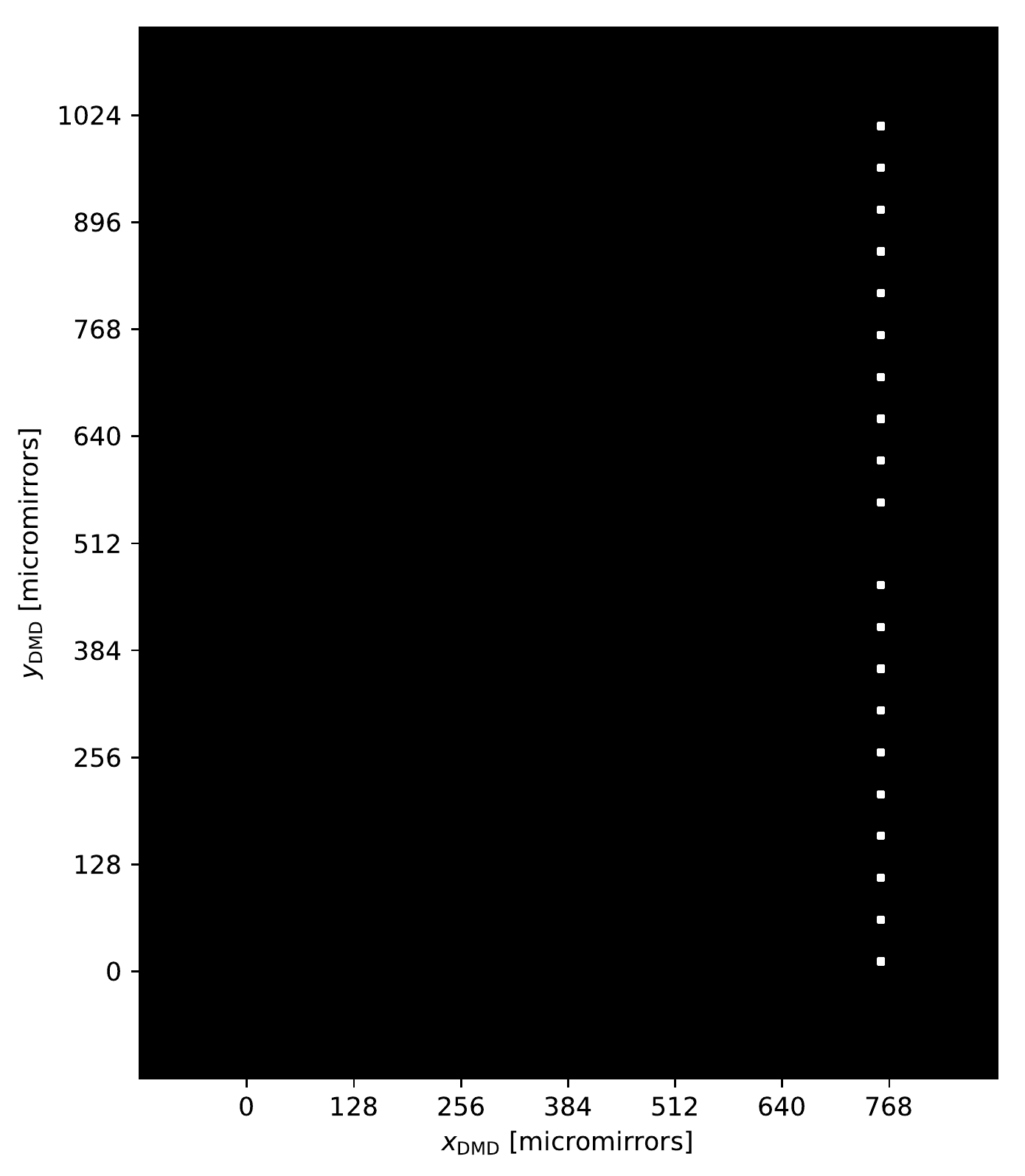}
        \caption{DMD with ON micromirrors along $x_\textrm{DMD}=758$}
        \label{fig:DMD-MOS_Footprint_DMD874}
    \end{subfigure}
    \begin{subfigure}{0.49\textwidth}
        \centering
        \includegraphics[width=0.9\textwidth]{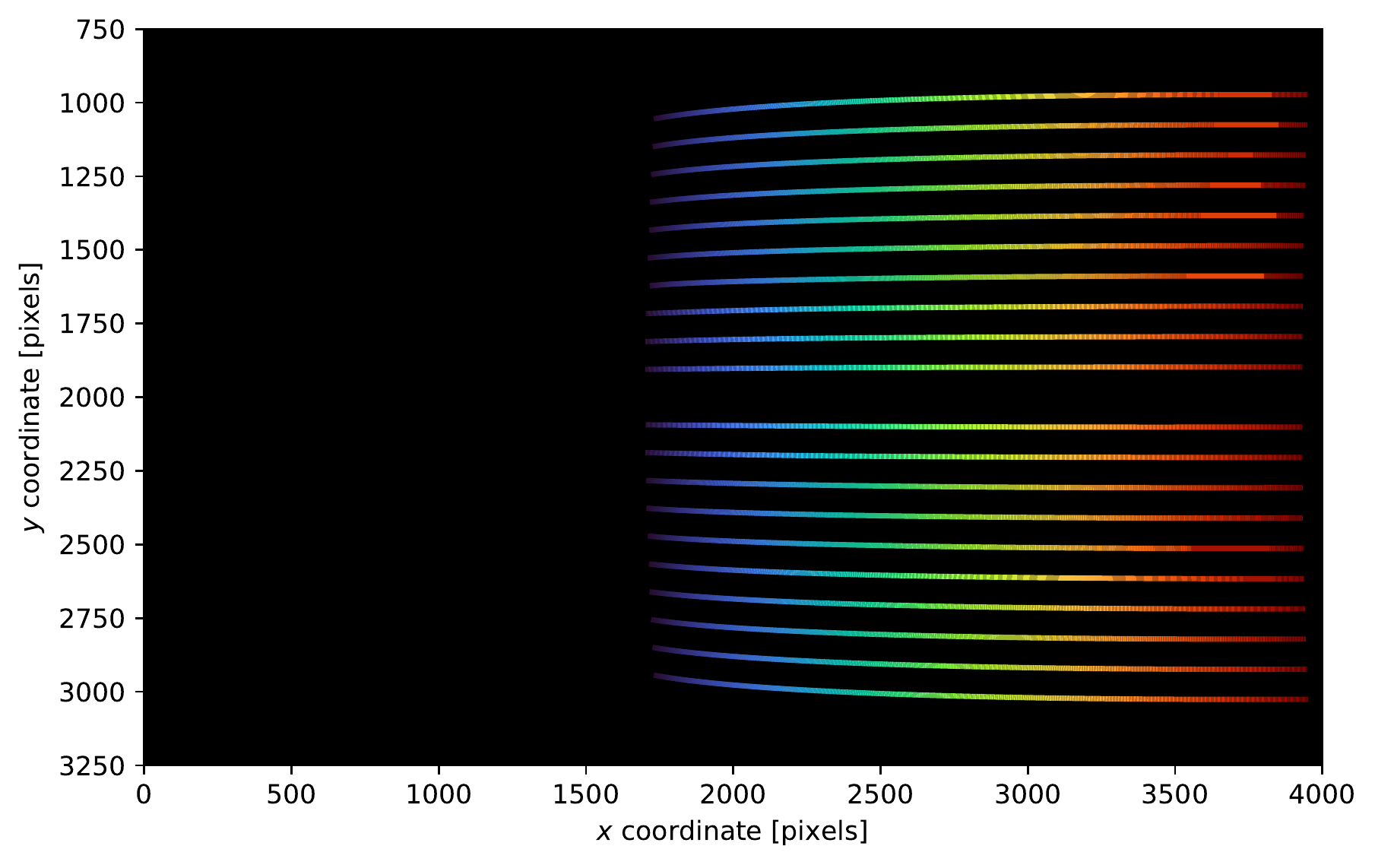}
        \caption{Corresponding spectra on detector for $x_\textrm{DMD}=758$}
        \label{fig:DMD-MOS_Footprint_Det874}
    \end{subfigure}
    \caption{Examples of some marked slit configurations for the DMD micromirrors (left), and simulated data for the resulting spectra on the detector (right). The ON micromirrors are shown as white squares in the left figures, and have been enlarged for visibility. For each fixed $x_\textrm{DMD}$, the DMD was programmed such that $N_\textrm{ON}=20$ ON micromirrors were periodically distributed along the $y_\textrm{DMD}$ direction.}
    \label{fig:DMD-MOS_Footprint}
\end{figure}

Each micromirror on the DMD can be identified by a unique coordinate pair $(x_\textrm{DMD}, y_\textrm{DMD})$, where $x_\textrm{DMD} \in [0,W_\textrm{DMD}-1]$ and $y_\textrm{DMD} \in [0,H_\textrm{DMD}-1]$. The DMD is oriented such that dispersion occurs along the $x_\textrm{DMD}$ direction. For calibration, the micromirrors were programmed such that there were $N_\textrm{ON}$ ON micromirrors periodically distributed along the $y_\textrm{DMD}$ direction for a certain fixed $x_\textrm{DMD}$ value. Simulated data was then acquired in Zemax OpticStudio for this DMD micromirror configuration, and this process was repeated for various other selected $x_\textrm{DMD}$ values. Figures \ref{fig:DMD-MOS_Footprint_DMD125}, \ref{fig:DMD-MOS_Footprint_DMD524}, and \ref{fig:DMD-MOS_Footprint_DMD874} show the DMD with $N_\textrm{ON}=20$ ON micromirrors periodically distributed in the $y_\textrm{DMD}$ direction for $x_\textrm{DMD}=9$, $x_\textrm{DMD}=408$, and $x_\textrm{DMD}=758$ respectively. Illustrations of the resulting spectra on the detector from these micromirror configurations are shown in Figures \ref{fig:DMD-MOS_Footprint_Det125}, \ref{fig:DMD-MOS_Footprint_Det524}, and \ref{fig:DMD-MOS_Footprint_Det874} respectively.

\subsection{About the DMD-MOS Simulated Data}

The resulting spectra of our DMD-MOS were simulated in Zemax OpticStudio using the Geometric Image Analysis ray tracing feature. The DMD was represented as an image with square holes at the image plane of the telescope, and the locations of the holes were changed for different micromirror configurations (for ON micromirrors at different $x_\textrm{DMD}$ values). Then for each micromirror configuration, $\sim$10000 rays were launched into the system at $K=13$ different wavelengths ranging from 400 nm to 700 nm. Rays on the detector that were from the same micromirror and of the same wavelength were then grouped together, and their centroid coordinate was computed. The positions of these centroids were then taken as the positions of the spectral peaks, forming the set $S$. Here, the set $S$ contains $N = N_\textrm{DMD}K = 20 \times 13 = 260$ points. In the future, for real-world data from the DMD-MOS, Gaussian functions could be fitted to slices of the spectrum image, as in Section \ref{sec:spec_lines_id}. Alternatively, the method in Section \ref{sec:spec_lines_id} could be used to roughly estimate the positions of spectral peaks; then for each spectral peak position corresponding to a specific wavelength and specific micromirror, a two-dimensional Gaussian function could be fitted to that spectral peak \cite{Hong17}.

Figures \ref{fig:DMD-MOS_125_S} and \ref{fig:DMD-MOS_874_S} show the set $S$ for the DMD configuration in Figures \ref{fig:DMD-MOS_Footprint_DMD125} and \ref{fig:DMD-MOS_Footprint_DMD874} respectively, with ON micromirrors along ${x_\textrm{DMD}=9}$ and ${x_\textrm{DMD}=758}$ respectively. From this arrangement of points, we can see that the points could be sparse in the spatial direction ($y$ direction on the detector), and that grouping them into parabola-shaped clusters may not be too obvious. This is why the $K$-parabolas algorithm is suitable for this problem. However, we realized that in the old design of our DMD-MOS, smile decreased with longer wavelengths, which is opposite of Equation (\ref{eq:delta_beta}). This behaviour may be due to other aberrations in the system, but more investigation is required to understand the causes. Nevertheless, the \mbox{$K$-parabolas} algorithm was still usable, only requiring minor modifications to the constraints in Section \ref{sec:problem}. The $a_i < a_{i+1} \:\forall i\in[1,K-1]$ constraint was instead changed to $a_i > a_{i+1} \:\forall i\in[1,K-1]$, and the $K$-parabolas algorithm performed well.


\subsection{Results for some DMD-MOS Simulated Data}

\begin{figure}[H]
    \centering
    \begin{subfigure}{0.49\textwidth}
        \centering
        \includegraphics[width=0.808\textwidth]{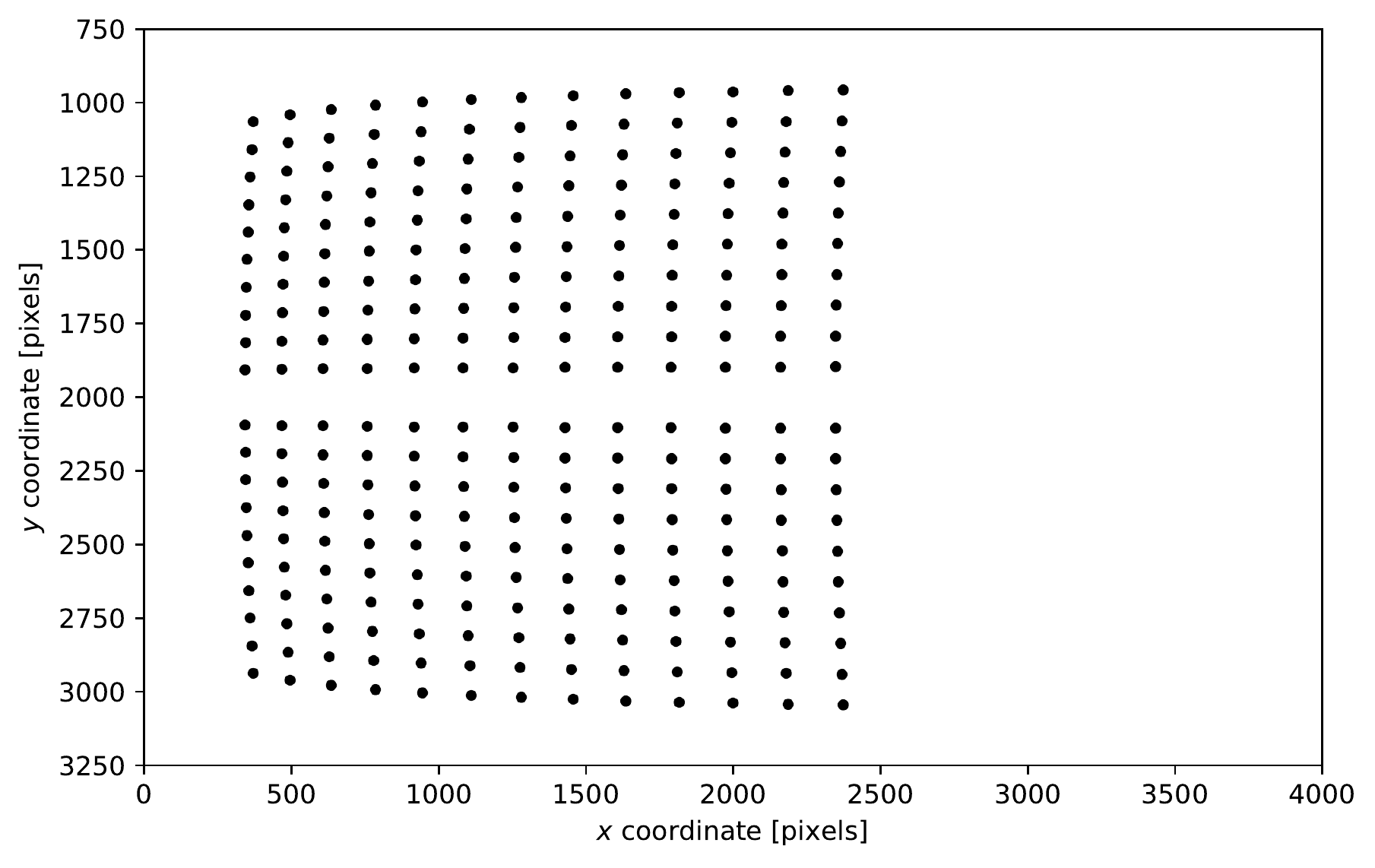}
        \caption{Simulated peak points $S$ from Zemax OpticStudio}
        \label{fig:DMD-MOS_125_S}
    \end{subfigure}
    \begin{subfigure}{0.49\textwidth}
        \centering
        \includegraphics[width=0.808\textwidth]{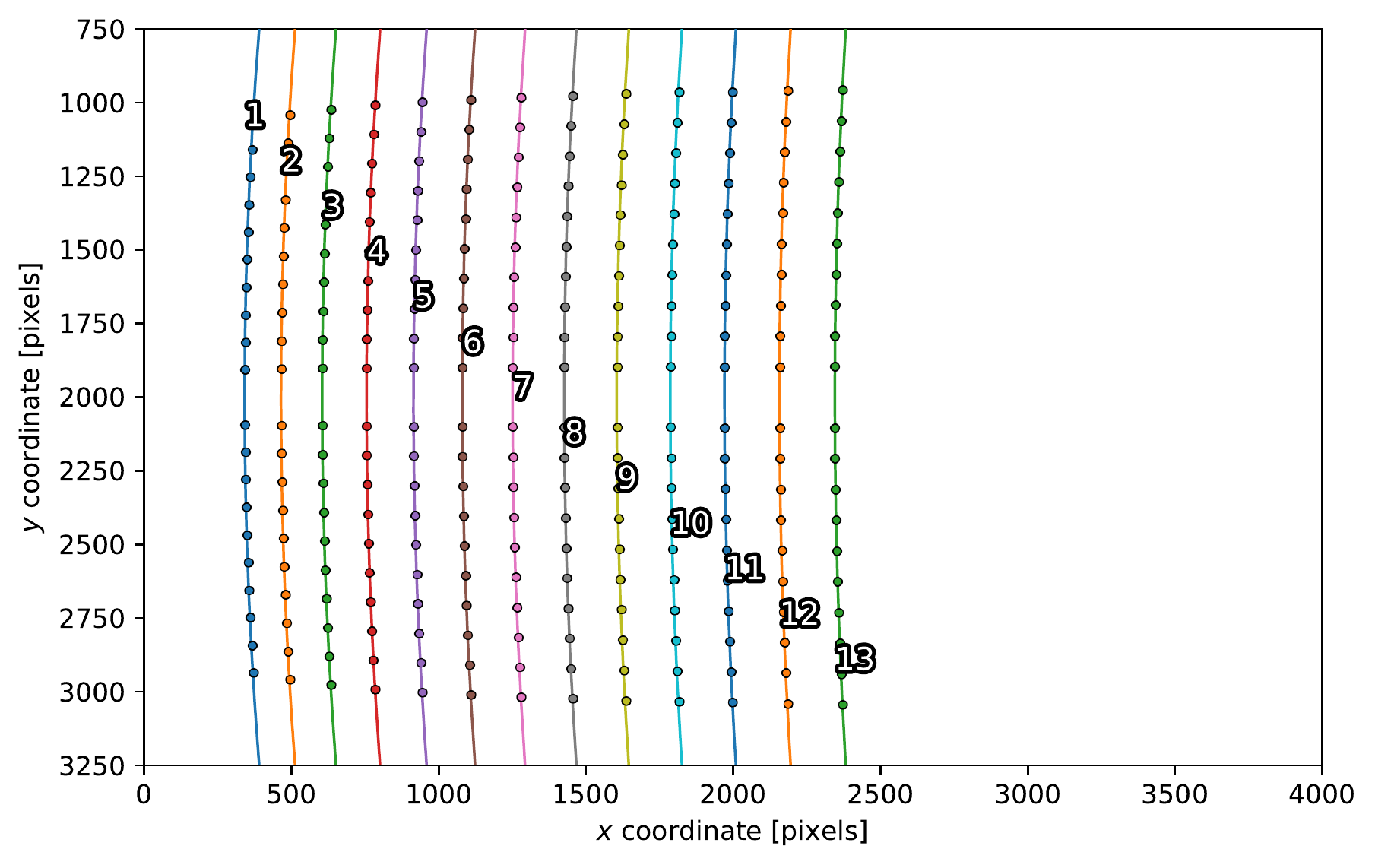}
        \caption{Result of $K$-parabolas and refined parabolas}
        \label{fig:DMD-MOS_125_clusters}
    \end{subfigure}
    \begin{subfigure}{0.49\textwidth}
        \centering
        \includegraphics[width=0.94\textwidth]{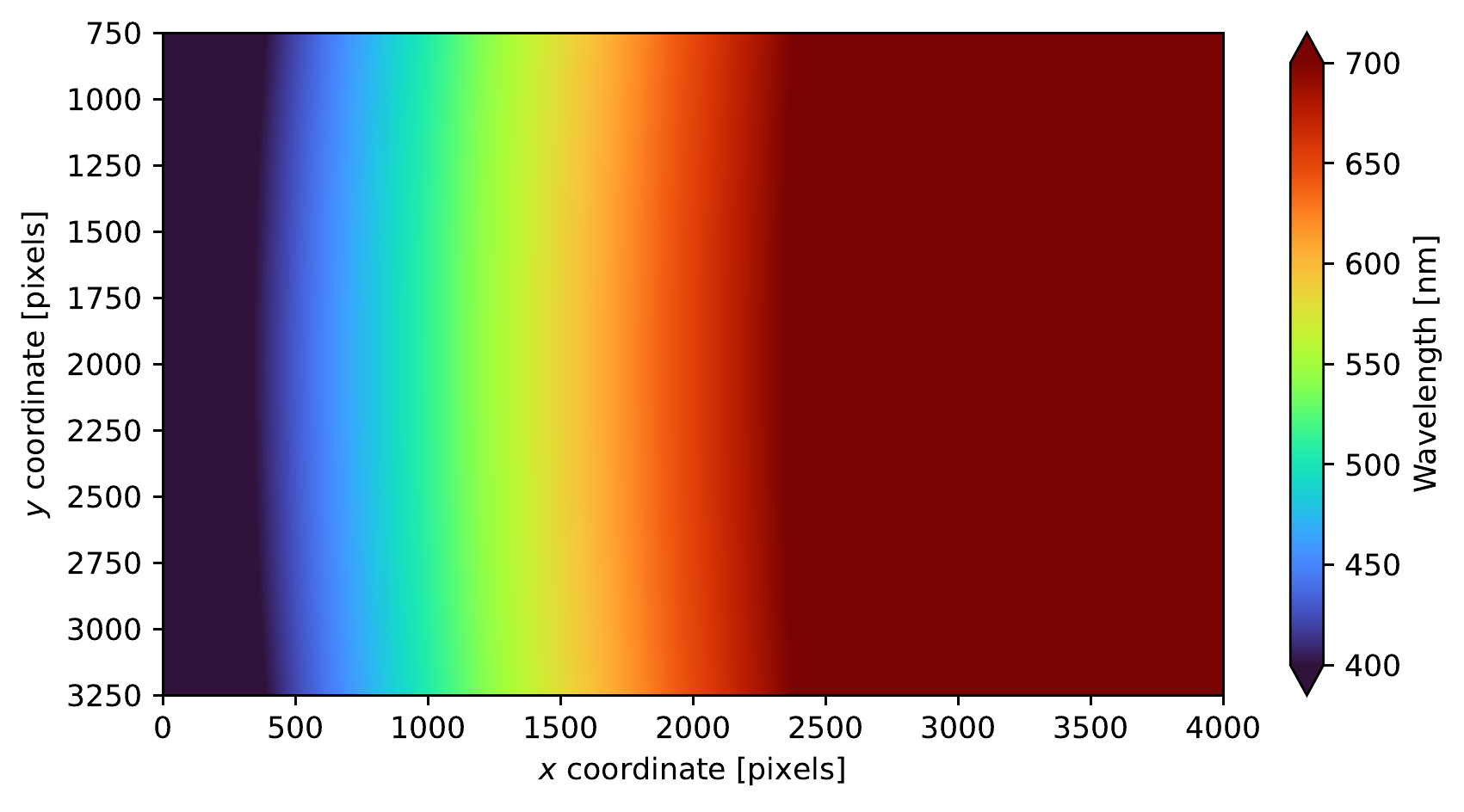}
        \caption{Pixel-to-wavelength mapping}
    \end{subfigure}
    \begin{subfigure}{0.49\textwidth}
        \centering
        \includegraphics[width=0.94\textwidth]{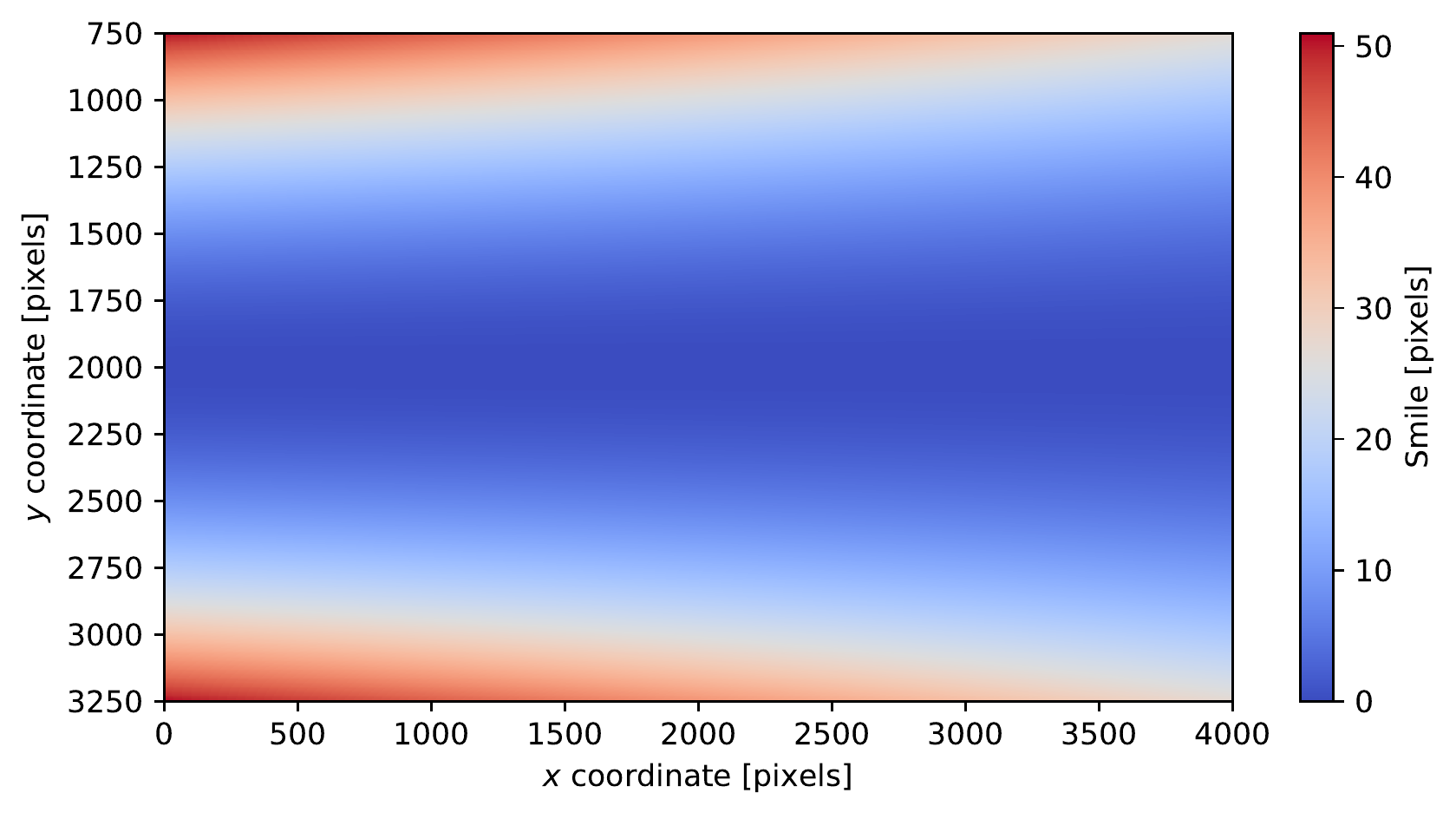}
        \caption{Amount of smile}
    \end{subfigure}
    \caption{Results for DMD-MOS simulated data, for the DMD configuration in Figure \ref{fig:DMD-MOS_Footprint_DMD125}, with ON micromirrors along $x_\textrm{DMD}=9$. Here, $K=13$.}
    \label{fig:DMD-MOS_125}
\end{figure}

\begin{figure}[H]
    \centering
    \begin{subfigure}{0.49\textwidth}
        \centering
        \includegraphics[width=0.808\textwidth]{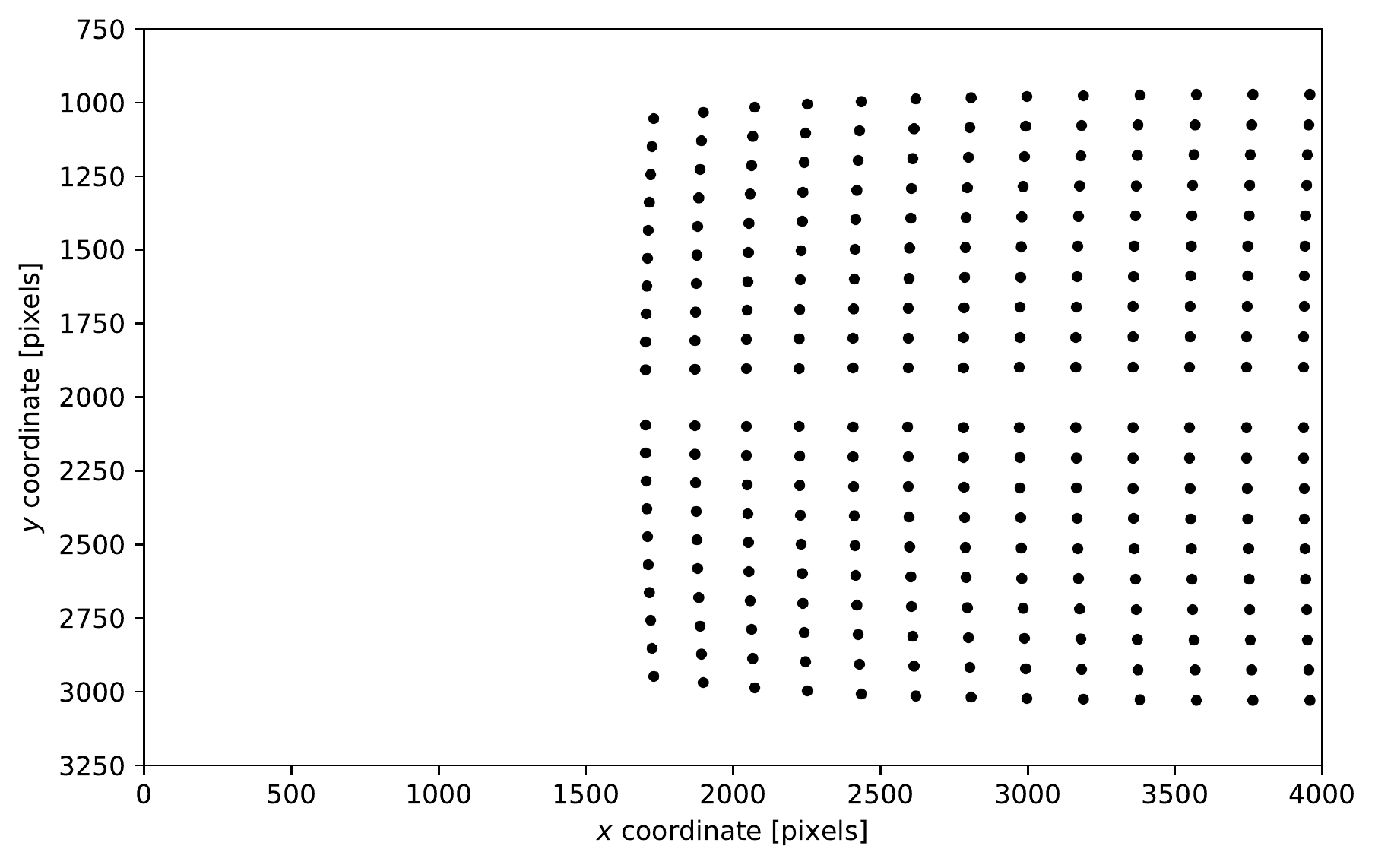}
        \caption{Simulated peak points $S$ from Zemax OpticStudio}
        \label{fig:DMD-MOS_874_S}
    \end{subfigure}
    \begin{subfigure}{0.49\textwidth}
        \centering
        \includegraphics[width=0.808\textwidth]{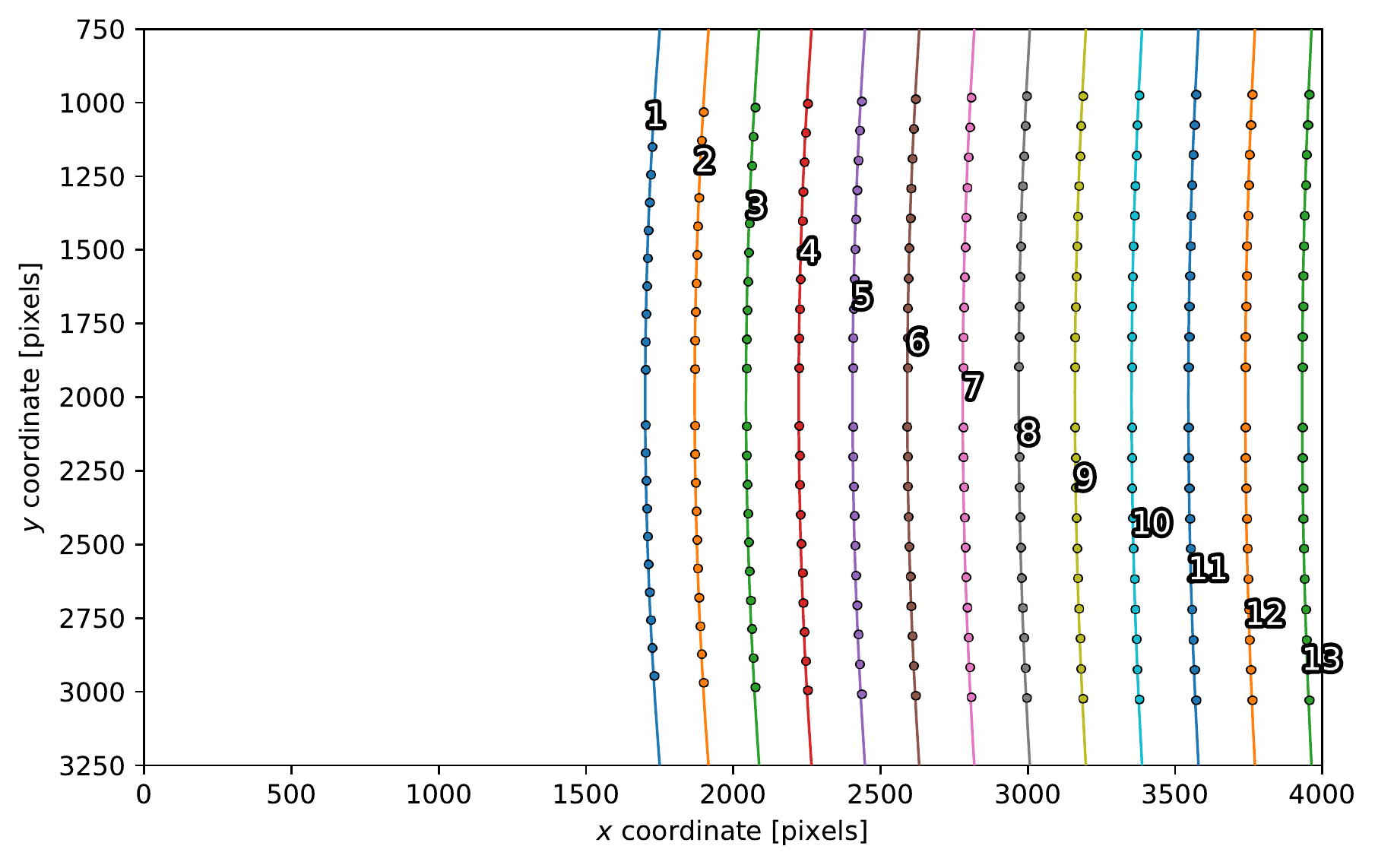}
        \caption{Result of $K$-parabolas and refined parabolas}
        \label{fig:DMD-MOS_874_clusters}
    \end{subfigure}
    \begin{subfigure}{0.49\textwidth}
        \centering
        \includegraphics[width=0.94\textwidth]{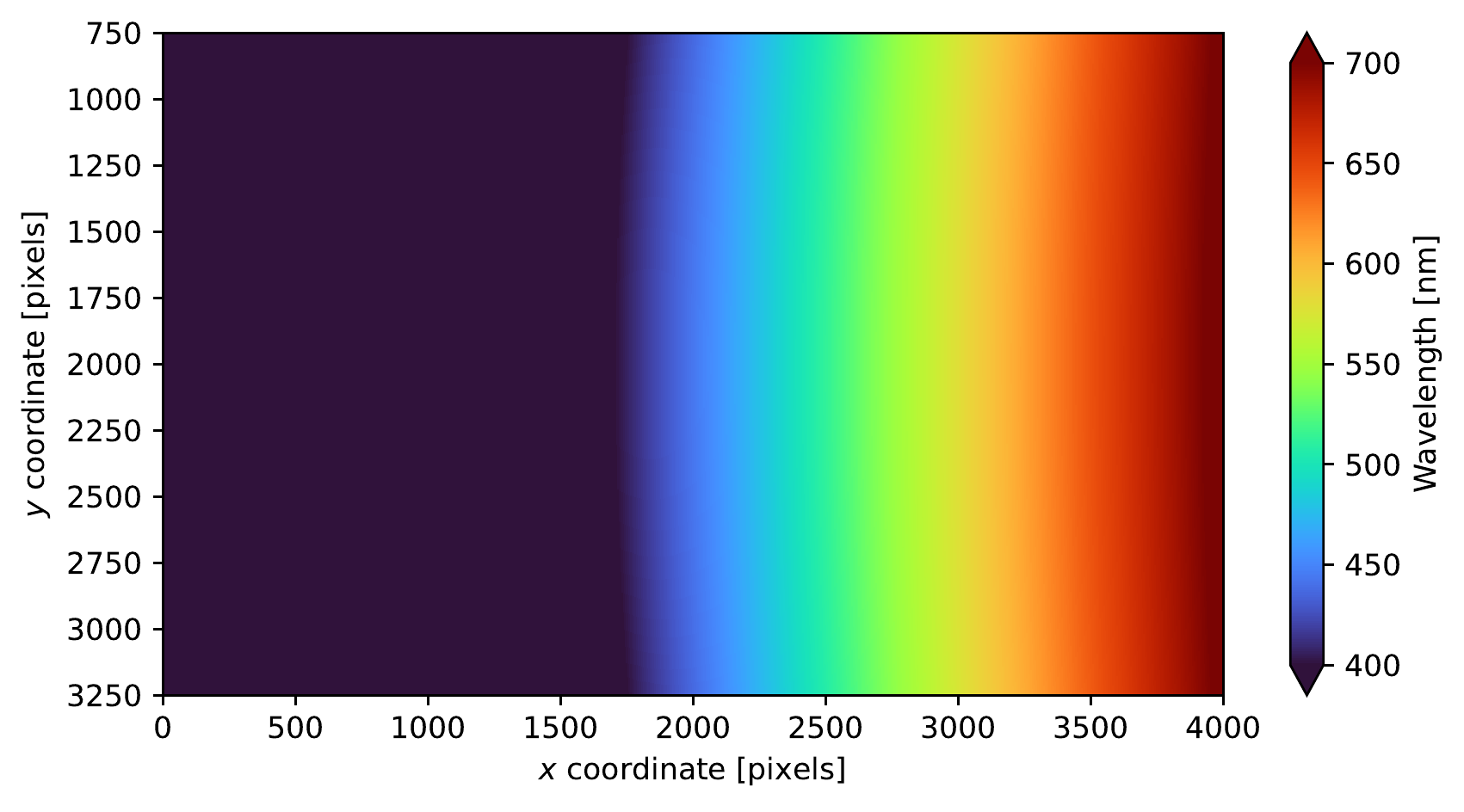}
        \caption{Pixel-to-wavelength mapping}
    \end{subfigure}
    \begin{subfigure}{0.49\textwidth}
        \centering
        \includegraphics[width=0.94\textwidth]{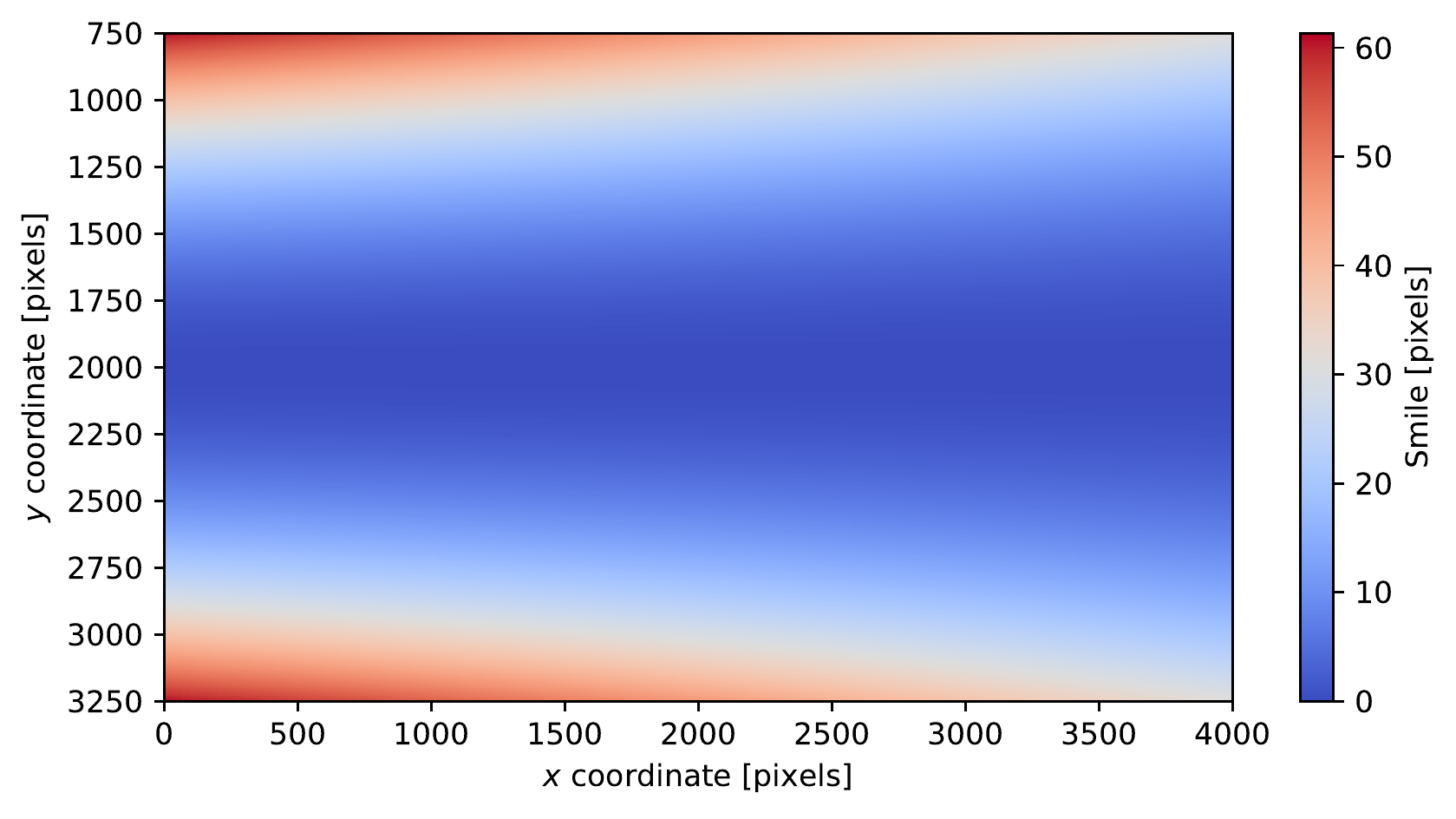}
        \caption{Amount of smile}
    \end{subfigure}
    \caption{Results for DMD-MOS simulated data, for the DMD configuration in Figure \ref{fig:DMD-MOS_Footprint_DMD874}, with ON micromirrors along $x_\textrm{DMD}=758$. Here, $K=13$.}
    \label{fig:DMD-MOS_874}
\end{figure}

Figures \ref{fig:DMD-MOS_125} and \ref{fig:DMD-MOS_874} show the final results for the DMD-MOS simulated data with the micromirrors configured as in Figures \ref{fig:DMD-MOS_Footprint_DMD125} and  \ref{fig:DMD-MOS_Footprint_DMD874} respectively. We can see that the spectra span 400 nm to 700 nm, which is consistent with the design of the DMD-MOS. For each wavelength, $K$-parabolas was able to find the equation of the parabola which models the corresponding spectral line, with small error. The final MAE of all the refined parabolas was 0.109 pixels and 0.113 pixels for Figures \ref{fig:DMD-MOS_125_clusters} and \ref{fig:DMD-MOS_874_clusters} respectively. The MAE would be even smaller if a larger number of rays were launched into the system in the simulation, so that the spectral peak positions would be more accurate.

\subsection{Keystone Correction}\label{sec:DMD-MOS_keystone}

\begin{figure}[H]
    \centering
    \includegraphics[width=0.395\textwidth]{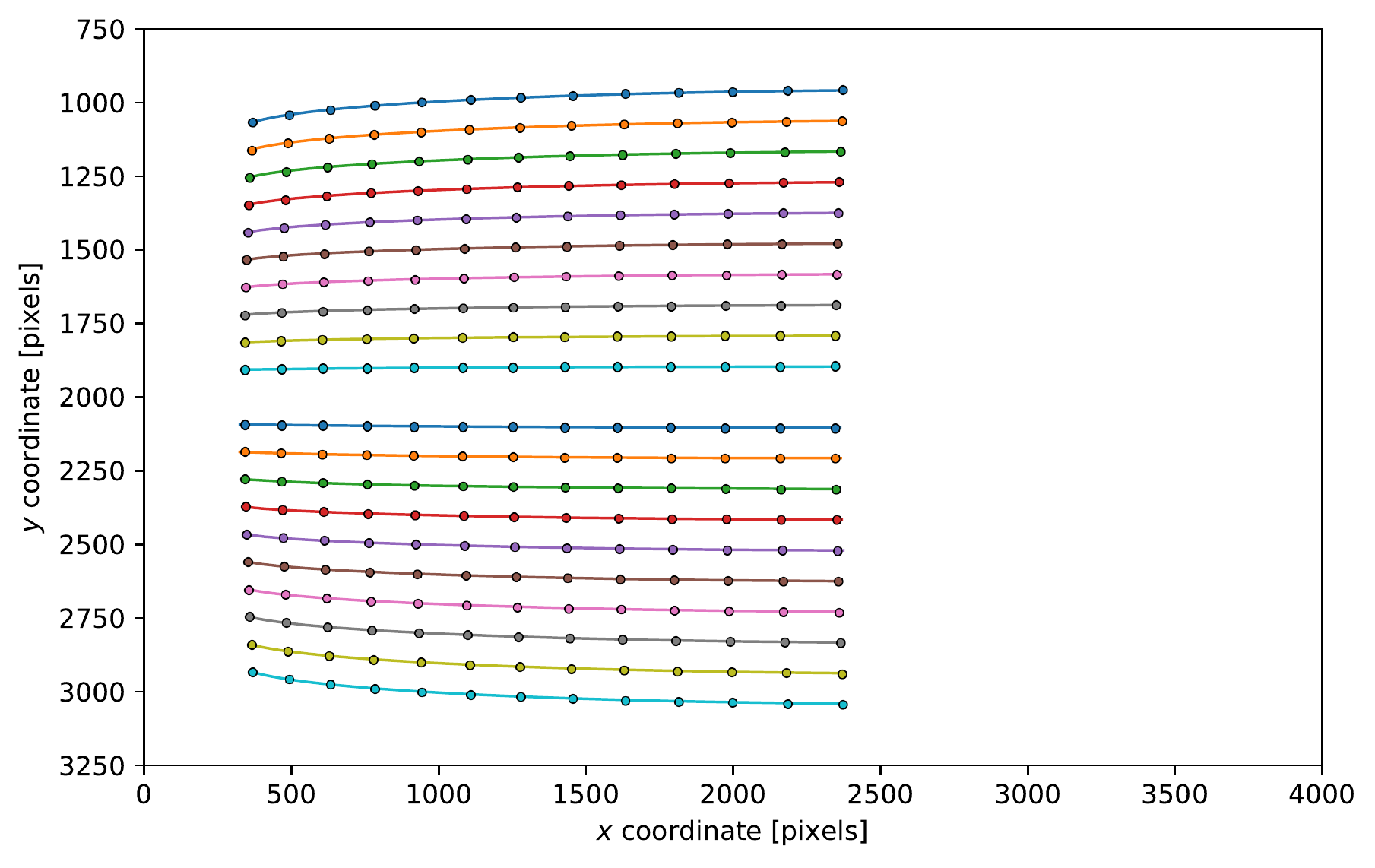}
    \caption{Keystone fits to the points $S$ in Figure \ref{fig:DMD-MOS_125_S}. For each group of points which corresponded to the same micromirror, and ellipse was fitted to that group.}
    \label{fig:DMD-MOS_125_keystone}
\end{figure}

With the marked slit configuration, keystone could also be corrected. Consider the set $S$ in Figure \ref{fig:DMD-MOS_125_S}, where the ON micromirrors were along $x_\textrm{DMD}=9$. To correct for keystone, firstly, for each ON micromirror, all points in $S$ which corresponded to that micromirror were grouped together. There are $N_\textrm{ON}$ of these groups. These groups are relatively easy to obtain, because we should have $|C_i|=N_\textrm{ON} \:\forall i\in[1,K]$ due to the marked slit. For each cluster $C_i$, $i\in[1,K]$, the $n$th point in the cluster (when we count moving downwards in the spectrum image) should belong to the $n$th micromirror, for $n\in[1,N_\textrm{ON}]$. Once these groups were obtained, an ellipse was fitted to each group. We found that an ellipse represented the points in each group well, which suggests that there may be some other geometric aberrations in the DMD-MOS, since a straight line could not be fitted to the points in each group. The ellipse fits for Figure \ref{fig:DMD-MOS_125_S} are shown in Figure \ref{fig:DMD-MOS_125_keystone}. More investigation is required in order to relate ellipse parameters with the micromirror $y_\textrm{DMD}$ positions, and this would be a next step.

\subsection{Trends Across the Detector}\label{sec:DMD-MOS_trends_detector}

\begin{figure}[H]
    \centering
    \begin{subfigure}{0.49\textwidth}
        \centering
        \includegraphics[width=\textwidth]{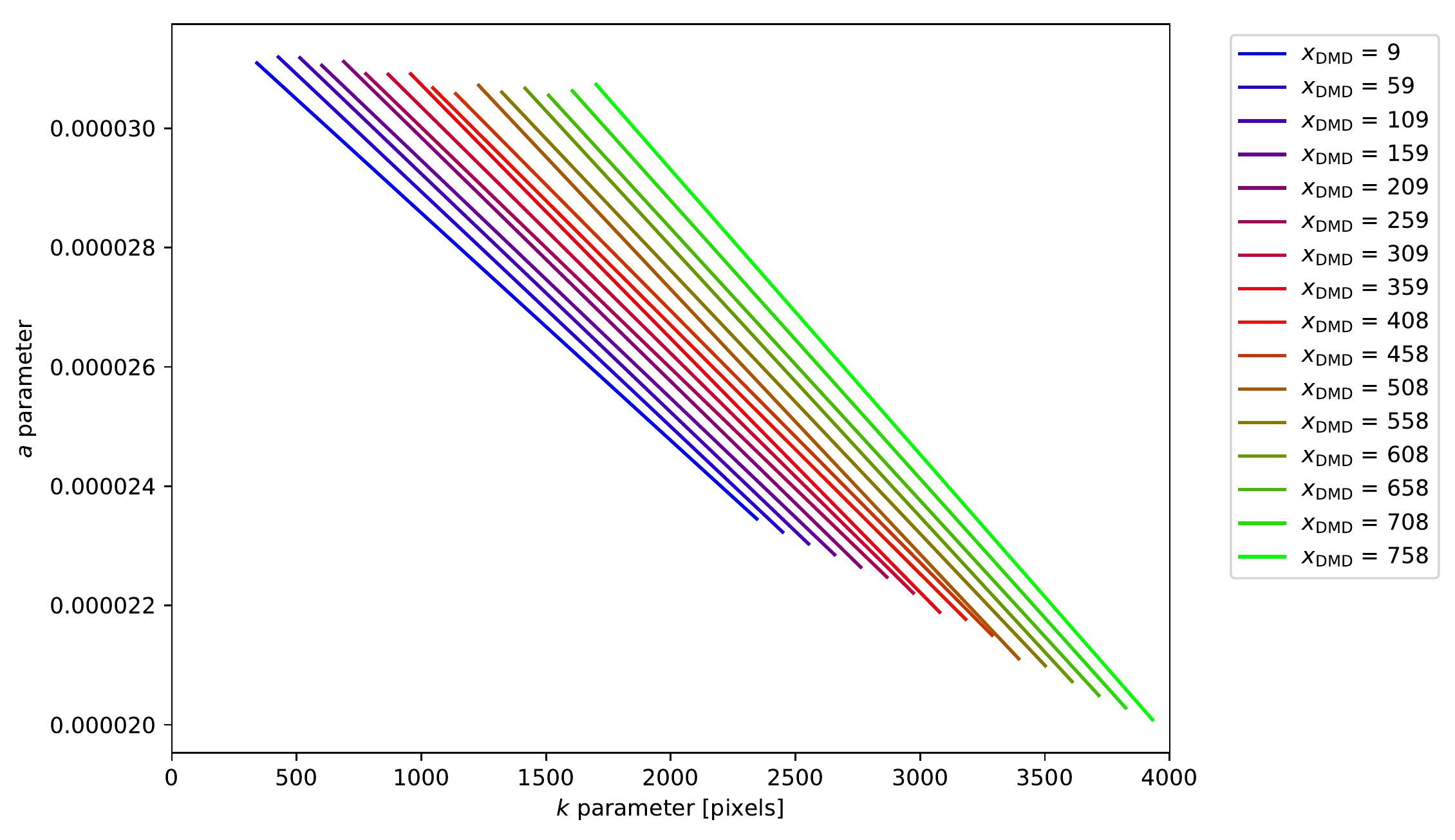}
        \caption{$a$ parameter for various $x_\textrm{DMD}$}
        \label{fig:DMD-MOS_a}
    \end{subfigure}
    \begin{subfigure}{0.49\textwidth}
        \centering
        \includegraphics[width=0.95\textwidth]{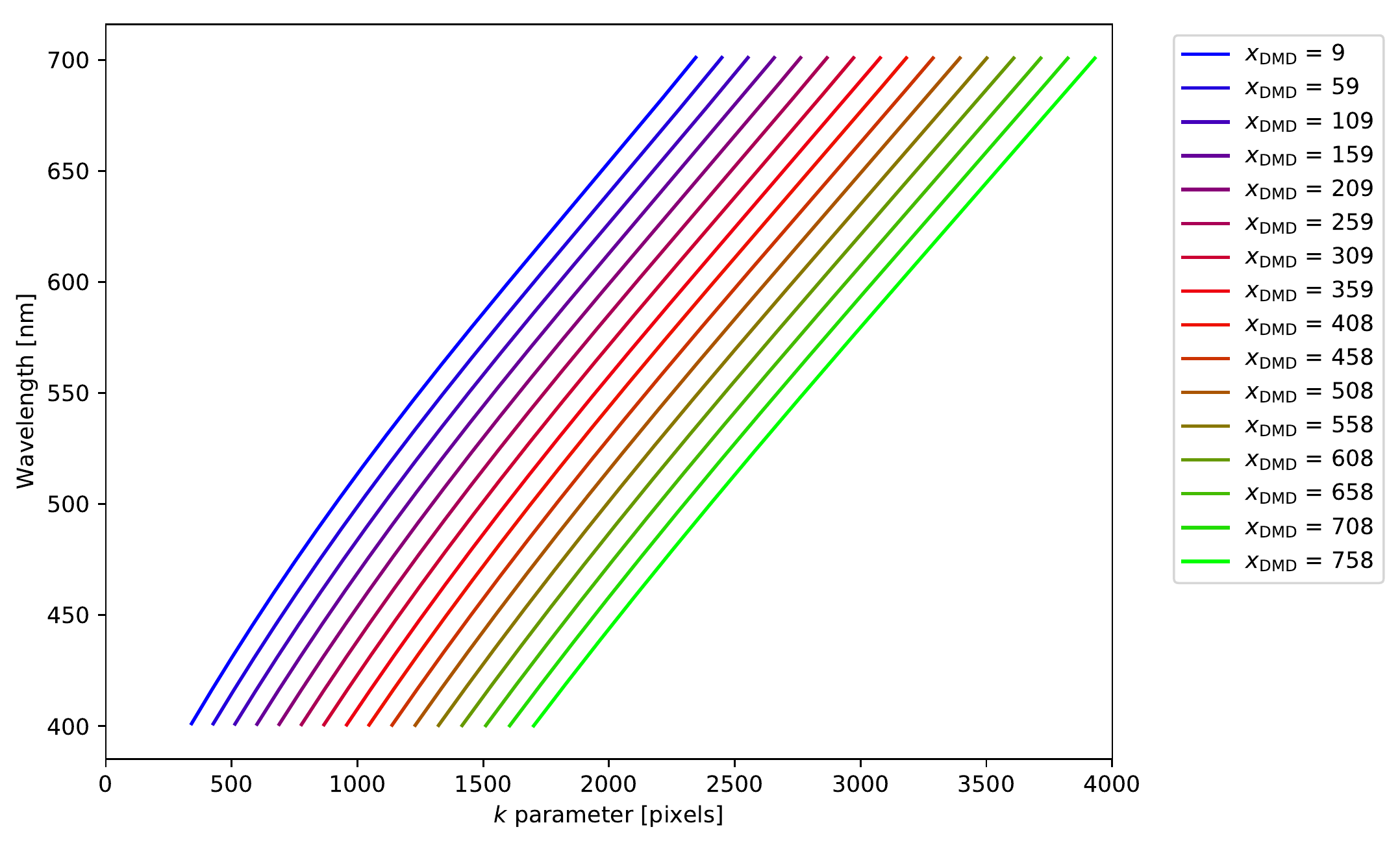}
        \caption{Wavelength $\lambda$ for various $x_\textrm{DMD}$}
        \label{fig:DMD-MOS_lambda}
    \end{subfigure}
    \caption{The bijective functions $a(k)$ and $\lambda(k)$ for some selected ON micromirror configurations.}
    \label{fig:DMD-MOS_a_lambda}
\end{figure}

Ultimately, we would like to develop a general relationship so that a specific micromirror and wavelength can be mapped to a specific pixel on the detector. To do this, certain selected ON micromirror configurations were used, with ON micromirrors at different selected $x_\textrm{DMD}$ values. Figures \ref{fig:DMD-MOS_a} and \ref{fig:DMD-MOS_lambda} respectively show the bijective functions $a(k)$ and $\lambda(k)$ (see Section \ref{sec:parameter_fits}) for a variety of selected ON micromirror configurations. Then to develop a general relationship, the next step would be to relate $a$ and $\lambda$ with $x_\textrm{DMD}$, and not just with $k$. Further investigation into this general relationship is required, and this is the larger problem. Nevertheless, the $K$-parabolas algorithm and pixel-to-wavelength mapping construction procedure are crucial steps in determining the different $a(k)$, $h(k)$, and $\lambda(k)$ functions which will ultimately be used to develop this general relationship.

There is also a more ad hoc but simple way to relate the DMD micromirror $x_\textrm{DMD}$ positions to the detector pixel positions. Note that in order to obtain the $a(k)$ and $\lambda(k)$ functions for each selected $x_\textrm{DMD}$ in Figure \ref{fig:DMD-MOS_a_lambda}, the simulated data for each selected ON micromirror configuration was being treated as an individual calibration problem. Technically, we could extend this and acquire simulated data for every ON micromirror configuration (for every $x_\textrm{DMD}\in [0,W_\textrm{DMD}-1]$, rather than just for some selected $x_\textrm{DMD}$ values), and then obtain a pixel-to-wavelength mapping for every ON micromirror configuration. Then the relationships between DMD micromirror $x_\textrm{DMD}$ positions and the detector pixel positions can be represented as a lookup table. However, this method may not be ideal as $W_\textrm{DMD}$ calibration spectra would have to be acquired, one for every ON micromirror configuration. Hence, the next step would be to try to develop a general relationship if possible.

\section{Next Steps}\label{sec:next_steps}

The $K$-parabolas algorithm can successfully group peak points into disjoint parabola-shaped clusters, allowing for curved spectral lines to be identified in both real-world data and simulated data. The four steps of $K$-parabolas discussed in Section \ref{sec:k_parabolas} work well, but the specifics of how these steps are implemented could be improved upon. For example, computing the orthogonal distances in Section \ref{sec:step2} is a bottleneck, and other improvements are possible, as have been proposed for the $K$-means algorithm when computing distances between points and centroids. Furthermore, the procedure of recomputing the parabola parameters in Section \ref{sec:step3} could also be improved upon. In addition, as touched upon in Section \ref{sec:remove_outliers}, the accuracy of $K$-parabolas depends on if the Gaussian fitting in Section \ref{sec:spec_lines_id} produced good peak points. More robust methods for finding the positions of spectral peaks could be implemented. Implementing these improvements would be some next steps in this research.

The method described in this paper has allowed for smile to be measured in a spectrum image. Keystone distortion can also be measured if a marked slit configuration is used, as was mentioned in Section \ref{sec:DMD-MOS_keystone} for the DMD-MOS. For the DMD-MOS, spectral peak positions corresponding to the same micromirror (same spatial coordinate) can be represented by an ellipse. A next step would be to relate the ellipse parameters with the spatial coordinate in order to develop a general relationship between keystone and detector pixel position. Finally, another next step would be to develop a general relationship between DMD micromirror position, wavelength, and detector pixel position, as mentioned in Section \ref{sec:DMD-MOS_trends_detector}.


\section{Conclusions}

We have demonstrated a novel method to accurately measure hyperspectral imaging distortion in grating spectrographs, using a clustering algorithm we developed called $K$-parabolas. This four-step algorithm iteratively groups spectral peak points into disjoint parabola-shaped clusters, allowing for the equation representing each spectral line in the image to be obtained automatically. With this information, a pixel-to-wavelength mapping can be obtained, and the smile for each pixel can be measured. Keystone measurement can be performed using a marked slit configuration. Our method has been verified on real-world data from a long-slit grating spectrograph and on simulated data from a DMD-based MOS. The $K$-parabolas algorithm is an important step in a larger problem to calibrate our DMD-MOS.

Some improvements could be made to our method. The accuracy of the Gaussian fitting process, which determines the spectral peak points, influences the results of $K$-parabolas. Hence, this process could be made more robust. In addition, while the four steps of $K$-parabolas work well, the specifics of how these steps are implemented could be improved upon, as is a similar case to $K$-means. Nevertheless, for each spectral line, our method has accurately found the equation of the parabola representing it. This information, in and of itself, is valuable and can be utilized for hyperspectral imaging distortion correction using a variety of methods.


\appendix

\section{Specifications of our Long-Slit Grating Spectrograph}\label{sec:appendix_spec}

\begin{figure}[H]
    \centering
    \begin{subfigure}{0.64\textwidth}
        \centering
        \includegraphics[width=\textwidth]{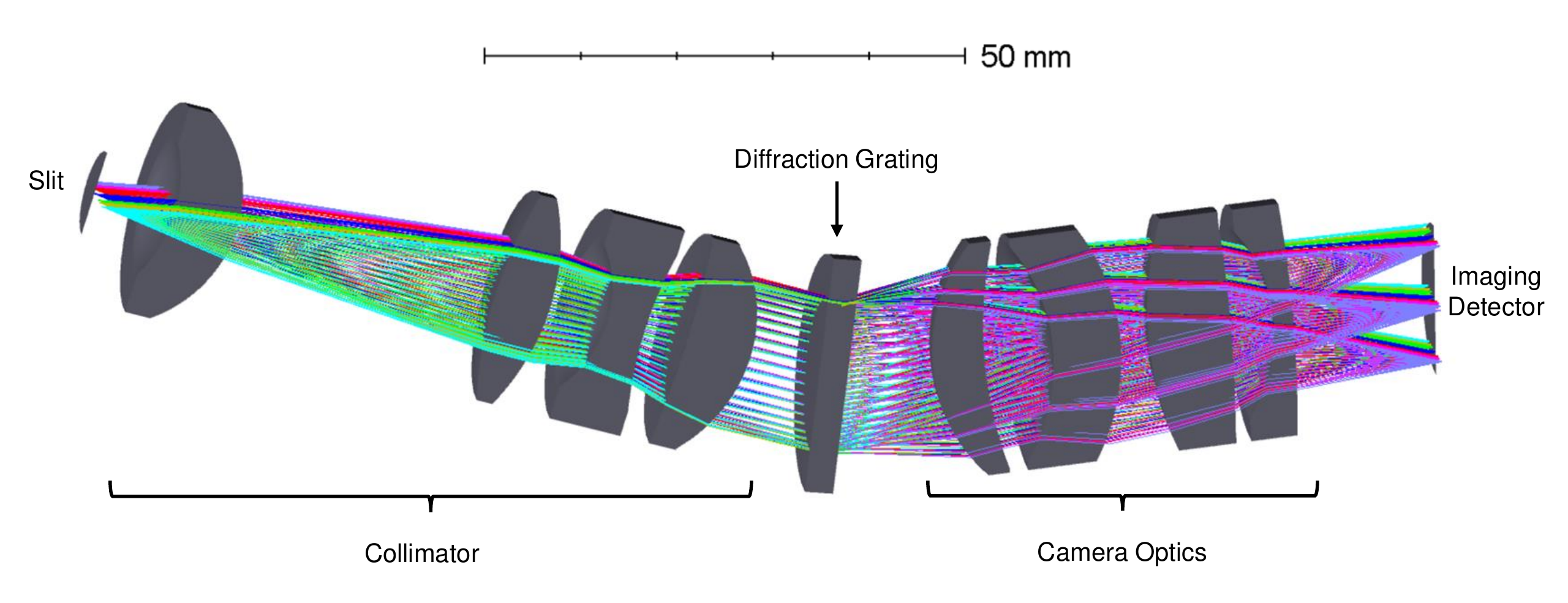}
        \caption{Zemax OpticStudio model}
        \label{fig:small_spec_Zemax}
    \end{subfigure}
    \begin{subfigure}{0.35\textwidth}
        \centering
        \includegraphics[width=\textwidth]{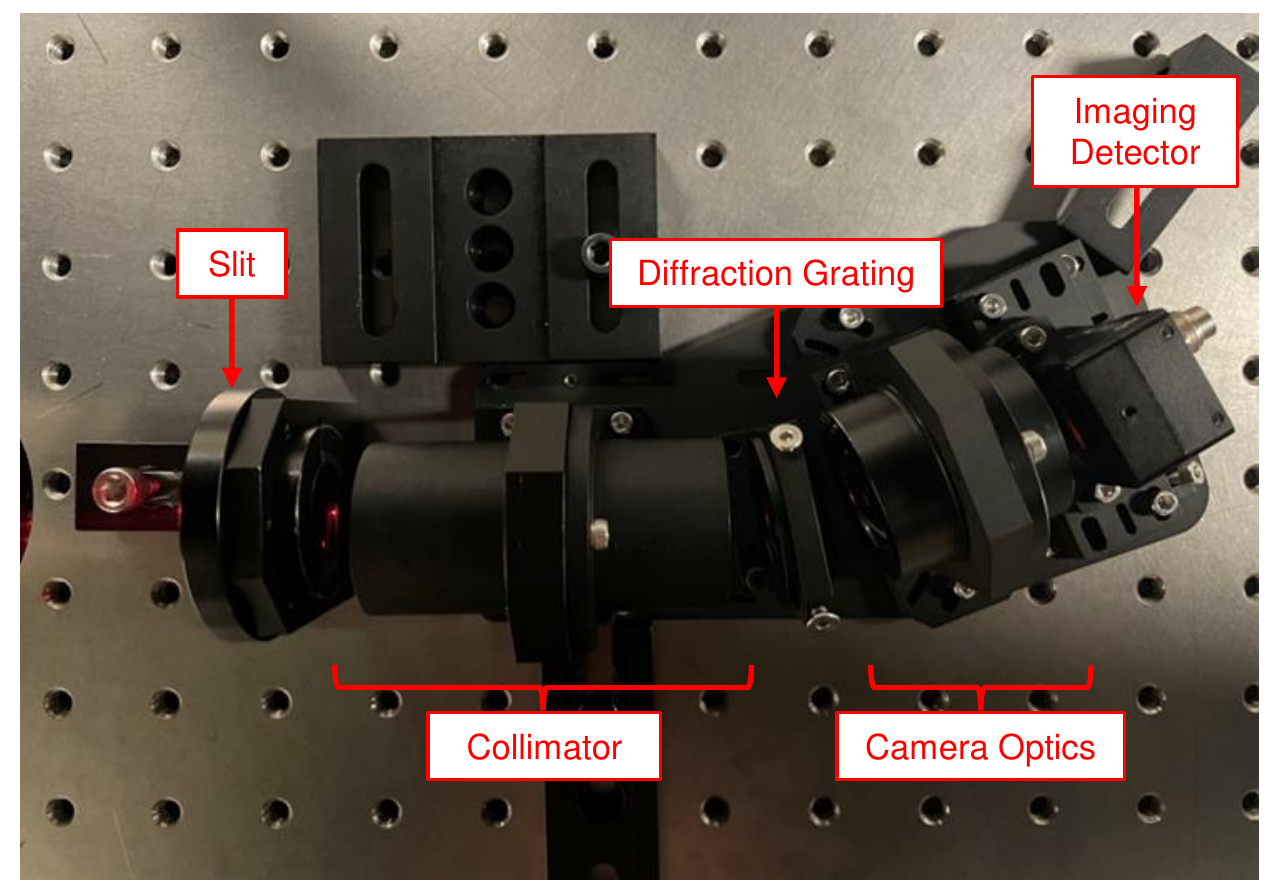}
        \caption{Actual spectrograph}
    \end{subfigure}
    \caption{Our long-slit grating spectrograph, from the slit to the imaging detector.}
    \label{fig:small_spec}
\end{figure}

Table \ref{table:small_spec_specs} lists some specifications of our long-slit grating spectrograph which was used to acquire the Hg calibration spectrum image in Figure \ref{fig:ex_spec_img}. Figure \ref{fig:small_spec} shows a Zemax OpticStudio model of this spectrograph and the actual spectrograph itself. This spectrograph was designed to be compact.

\begin{table}[H]
\begin{center}
    \begin{tabular}{|l|l|} 
    \hline
    Parameter or Component & Value or Details \\
    \hline
    \hline
    Spectral Range & 400 nm to 900 nm\\ \hline
    Spectral Resolution & $\sim$1000 at 550 nm\\ \hline
    Slit Width & 15 \textmu m\\ \hline
    Imaging Detector & Sony IMX174 (CMOS), $1936\times1216$, 5.86 \textmu m pixel pitch \\ \hline
    Imaging Performance & FWHM is within 3 pixels\\
    \hline
    \end{tabular}
    \caption{Some specifications of our long-slit grating spectrograph.}
    \label{table:small_spec_specs}
\end{center}
\end{table}

\section{Specifications of our DMD-based Multi-Object Spectrograph  (Old Design)}\label{sec:appendix_DMD-MOS}

\begin{figure}[H]
    \centering
    \includegraphics[width=0.9\textwidth]{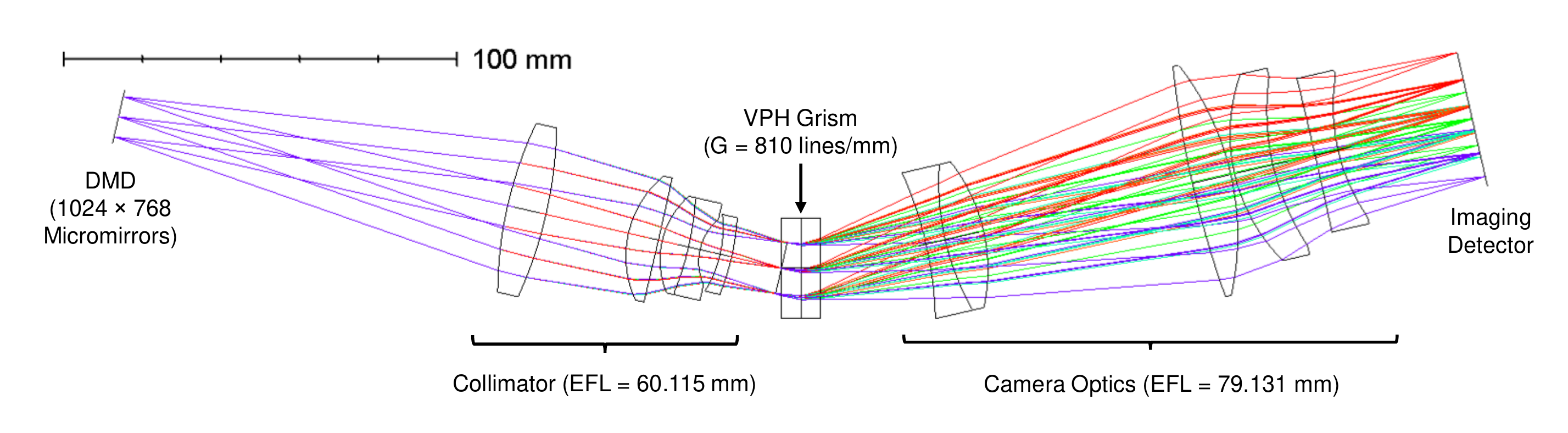}
    \caption{Zemax OpticStudio model of the old design of our DMD-MOS, from the DMD to the imaging detector.}
    \label{fig:DMD-MOS_Zemax}
\end{figure}

Table \ref{table:DMD-MOS_specs} lists some specifications of the old design of our DMD-MOS. Figure \ref{fig:DMD-MOS_Zemax} shows a model of this DMD-MOS in Zemax OpticStudio, which was used to generate the simulated data in Section \ref{sec:DMD-MOS}. Comparing Figure \ref{fig:DMD-MOS_Zemax} to Figure \ref{fig:small_spec_Zemax}, we see that the DMD acts as a programmable slit. Note that out of the $1024 \times 768$ DMD micromirrors, we only use $1000\times750$. The current design of our DMD-MOS is discussed in another paper published in these proceedings \cite{Chen2022}.

\begin{table}[H]
\begin{center}
    \begin{tabular}{|l|l|} 
    \hline
    Parameter or Component & Value or Details \\
    \hline
    \hline
    Spectral Range & 400 nm to 700 nm\\ \hline
    Spectral Resolution & $\sim$1000 at 550 nm\\ \hline
    DMD & Texas Instruments DLP7000, $1024 \times 768$, 13.7 \textmu m micromirror pitch \\ \hline
    Imaging Detector & Alta ASP-F16 (CCD), $4096\times4096$, 9 \textmu m pixel pitch \\
    \hline
    \end{tabular}
    \caption{Some specifications of the old design of our DMD-MOS.}
    \label{table:DMD-MOS_specs}
\end{center}
\end{table}

\acknowledgments 

M.C.H. Leung would like to thank the Dunlap Institute for Astronomy and Astrophysics for their support of this research in the form of a Summer Undergraduate Research Program (SURP) Fellowship.
 
\bibliography{report} 
\bibliographystyle{spiejour} 

\end{document}